\documentclass[12pt]{article}

\usepackage[mathscr]{eucal}
\usepackage{amsfonts}
\usepackage{amssymb}
\usepackage{amsthm}

\newcommand{\be}{\begin{equation}}
\newcommand{\ee}{\end{equation}}
\newcommand{\bea}{\begin{eqnarray}}
\newcommand{\eea}{\end{eqnarray}}

\begin{document}

\begin{titlepage}
\vspace*{1.5cm}

\renewcommand{\thefootnote}{\dag}
\begin{center}
{\LARGE\bf  The algebraic structure of}

\vspace{0.5cm}

{\LARGE\bf Galilean superconformal symmetries}

\vspace{1.5cm}
\renewcommand{\thefootnote}{\star}

{\large\bf Sergey Fedoruk}${}^{\,1}$,\,\,\, \,\,
{\large\bf Jerzy Lukierski}${}^{\,2}$ \vspace{1.5cm}

${}^{1)}${\it Bogoliubov  Laboratory of Theoretical Physics, JINR,}\\
{\it 141980 Dubna, Moscow region, Russia} \\
\vspace{0.1cm}

{\tt fedoruk@theor.jinr.ru}\\
\vspace{0.5cm}

${}^{2)}${\it Institute for Theoretical Physics, University of Wroc{\l}aw,}\\
{\it pl.
Maxa Borna 9, 50-204 Wroc{\l}aw, Poland} \\
\vspace{0.1cm}

{\tt lukier@ift.uni.wroc.pl}\\
\vspace{1cm} \setcounter{footnote}{0}

\end{center}
\vspace{0.2cm} \vskip 0.6truecm  \nopagebreak

\begin{abstract}
\noindent The semisimple part of $d$-dimensional Galilean conformal algebra $g^{(d)}$
is given by $h^{(d)}=O(2,1)\oplus O(d)$, which after adding via semidirect sum the $3d$-dimensional Abelian algebra
$t^{(d)}$ of translations, Galilean boosts and constant accelerations
completes the construction.
We obtain Galilean superconformal algebra $G^{(d)}$ by firstly defining the semisimple
superalgebra $H^{(d)}$ which supersymmetrizes $h^{(d)}$, and further by considering the expansion of $H^{(d)}$
by tensorial and spinorial graded Abelian charges in order to supersymmetrize the Abelian generators of $t^{(d)}$.
{}For $d\,{=}\,3$ the supersymmetrization of $h^{(3)}$ is linked with specific model of ${\cal N}{=}\,4$ extended superconformal mechanics,
which is described by the superalgebra $D(2,1;\alpha)$ if $\alpha\,{=}\,1$.
We shall present as well the alternative derivations of extended Galilean superconformal algebras
for $1\,{\leq}\,d\,{\leq}\,5$ by employing the {I}n\"{o}n\"{u}-Wigner contraction method.
\end{abstract}



\newpage

\end{titlepage}

\setcounter{equation}{0}
\section{Introduction}

\quad\, Many applications
of the relativistic AdS/CFT correspondence \cite{Mal,GuKlPol,Wit} brings
the question of its nonrelativistic limit.
For such purpose we should define the nonrelativistic conformal field theory and the algebra of nonrelativistic conformal symmetries.
If we recall that the relativistic AdS/CFT correspondence
has its best justification after the supersymmetric extension
(e.g. $D\,{=}\,5$ supergravity versus $N\,{=}\,4$ $D\,{=}\,4$ SYM theory)
it appears important to study as well the nonrelativistic superconformal symmetries.

In seventies \cite{Ja,Ha,Ni,BuPeSo} the Galilean symmetries were extended to the Schr\"{o}dinger symmetries,
by suplementing the Galilean Lie algebra by two additional generators: $\mathbf{D}$ (dilatations) and
$\mathbf{K}$ (extensions). Because of conformal nature of the generators $\mathbf{D}$, $\mathbf{K}$
the Schr\"{o}dinger symmetries were named as the nonrelativistic conformal symmetries. However
soon appeared better candidate for such symmetries, described by the Galilean conformal algebra (GCA):\footnote{
Galilean conformal algebra (GCA) was considered in \cite{Barut,HavPl,Henkel,NOR};
the case $D=2+1$, unique dimension permitting central charge, was considered in \cite{LSZ2}.
The GCA-invariant mechanics models with $d\,{\geq}\,1$ were studied recently in \cite{FIL-W}; for the
review of geometric techniques related with nonrelativistic conformal algebras see \cite{DuHo}.
}
the $c\,{\rightarrow}\,\infty$ contraction limit of the relativistic conformal
algebra \cite{Barut,HavPl,Henkel,NOR,LSZ2}. The Galilean conformal symmetries
describe massless nonrelativistic systems -- contrary to the case of Schr\"{o}dinger algebra
the central extension of Galilean conformal
algebras which introduces the nonrelativistic mass parameter is not allowed.
It was further realized (for recent review see \cite{DuHo}) that if we relate nonrelativistic
conformal symmetries with Newton-Cartan structure of nonrelativistic space and time,
one gets families of infinite-dimensional Virasoro-like conformal algebras (see e.g. \cite{Bag1,Bag2}).
In particular, there was introduced the notion of generalized Schr\"{o}dinger
algebras \cite{Henkel} which do depend on the numerical parameter $z$ called dynamical exponent,
characterizing the scaling properties of space and time coordinates.
If $z\,{=}\,1$ we obtain GCA, and for $z\,{=}\,2$ one gets the Schr\"{o}dinger algebra;
recently it has been shown  \cite{DuHo11} that for rational choices of $z\,{=}\,\frac{2}{N}$ ($N=1,2,...$)
all the generalized Schr\"{o}dinger algebras are finite-dimensional.

In this paper we restrict our studies to the Galilean conformal algebras and their
supersymmetric extensions.
Superalgebras of the Galilean superconformal symmetry has been constructed in \cite{AL,Sa} (see also \cite{GKTown,GGKam})
from the relativistic conformal superalgebras by the contraction methods.
In this paper we argue how the structure of the Galilean superconformal algebras (SUSY GCA)
is related with the superalgebras describing the models of extended superconformal mechanics.
We show that the supersymmetrization of the Abelian sector of translations, Galilean boosts and constant accelerations
requires the addition of suitable sector of graded Abelian fermionic supercharges
and possibly additional Abelian bosonic charges.

Let us consider the Galilean superconformal algebras as enlargement
of simple standard superalgebras by the bosonic and fermionic graded Abelian charges.\footnote{
For the enlargements of Lie (super)algebras by extension and expansion procedure see
e.g.  \cite{AIPV}.
}
Taking into account that the Galilean superconformal algebras  contain as its semisimple part the conformal algebra $O(2,1)$
of conformal mechanics \cite{AFF,IKL}
and the algebra of spatial rotations $O(d)$,\footnote
{We will denote the algebras by capital letters for convenience, making no distinction in the notation of the algebras and corresponding group.
This will not lead below to any misunderstanding.
}
we firstly introduce the simple superalgebras $G^{(d)}_{+}$
which contain as bosonic subalgebra $O(2,1)\oplus O(d)$.
Further we extend the superalgebra $G^{(d)}_{+}$ by graded Abelian superalgebra
$G^{(d)}_{\,-}=({\mathcal{B}}^{(d)}_{\,-},{\mathcal{Q}}^{(d)}_{\,-})$  where
\begin{equation}\label{S-}
G^{(d)}_{-}\,:\qquad [{\mathcal{B}}^{(d)}_{\,-},{\mathcal{B}}^{(d)}_{\,-}]=0\,,\qquad
[{\mathcal{B}}^{(d)}_{\,-},{\mathcal{Q}}^{(d)}_{\,-}]=0\,,\qquad \{{\mathcal{Q}}^{(d)}_{\,-},{\mathcal{Q}}^{(d)}_{\,-}\}=0\,.
\end{equation}

We recall that the conformal Galilean algebra $g^{(d)}$
is described as the following semidirect sum
\begin{equation}\label{g}
g^{(d)}\,:\qquad \left( O(2,1)\oplus O(d) \right)\subset\!\!\!\!\!\!+ \,{\mathscr{A}}^{(3d)}
\end{equation}
with the real generators $\mathbf{R}_{r}$
($r=0,1,2$), $\mathbf{J}_{ij}=-\mathbf{J}_{ji}$
($i,j=1,...,d$), $\mathbf{A}_{r,i}=(\mathbf{P}_{i},\mathbf{B}_{i},\mathbf{F}_{i})$ \footnote{
{}For clarity we denote the specific generators of the Galilean superalgebra by bold letters,
unlike other ones, e.g. defining the relativistic conformal superalgebra.
}
\begin{eqnarray}\label{o21}
O(2,1)&:&\qquad\left[\mathbf{R}_{r}, \mathbf{R}_{s}\right]=i\,\epsilon_{rs}{}^t \,\mathbf{R}_{t} \,,
\\
\label{0-d}
O(d)&:&\qquad \left[\mathbf{J}_{ij}, \mathbf{J}_{kl}\right]=
i\left(\delta_{ik}\,\mathbf{J}_{jl}+ \delta_{jl}\,\mathbf{J}_{ik}-\delta_{il}\,\mathbf{J}_{jk}-\delta_{jk}\,\mathbf{J}_{il}\right),
\\
\label{alg-com}
{\mathscr{A}}^{(3d)}&:&\qquad \left[\mathbf{A}_{r,i}, \mathbf{A}_{s,j}\right]=0
\end{eqnarray}
where we denote by $\mathbf{P}_{i}$ $d$-dimensional momenta, by $\mathbf{B}_{i}$
the Galilean boosts and by $\mathbf{F}_{i}$ the generators of constant accelerations.
The generators $\mathbf{A}_{r,i}$ satisfy the following covariance relations
\begin{equation}\label{o-c-c}
\left[\mathbf{R}_{r}, \mathbf{A}_{s,i}\right]=i\,\epsilon_{rs}{}^t \,\mathbf{A}_{t,i}\,, \qquad\quad
\left[\mathbf{J}_{ij}, \mathbf{A}_{r,k}\right]=i\left(\delta_{ik}\,\mathbf{A}_{r,j}-\delta_{jk}\,\mathbf{A}_{r,i}\right).
\end{equation}
We shall postulate that the generators $\mathbf{A}_{r,i}$ belong to $G^{(d)}_{-}$  ($\mathbf{A}_{r,i}\in {\mathcal{B}}^{(d)}_{\,-}$).
In our SUSY GCA which we shall denote
by $G^{(d)}$ the graded Abelian generators of $G^{(d)}_{-}=({\mathcal{B}}^{(d)}_{\,-},{\mathcal{Q}}^{(d)}_{\,-})$
satisfy (\ref{S-})  and
\begin{equation}\label{S+-}
[G^{(d)}_{+},G^{(d)}_{-}\}\subset G^{(d)}_{-} \,.
\end{equation}
We arrive therefore at the semidirect sum structure of SUSY GCA
\begin{equation}\label{S-gen}
G^{(d)} =G^{(d)}_{+} \subset\!\!\!\!\!\!+ \,\, G^{(d)}_{-}\,.
\end{equation}
It should be added that the structure (\ref{S-gen}) also follows from the {I}n\"{o}n\"{u}-Wigner (IW) supercoset contraction procedure
\cite{IW,GKTown,GGKam,Sa}. In this paper we shall show as well how to arrive at the structure
of SUSY GCA $G^{(d)}$ by suitable enlargement of the semisimple superalgebra $G^{(d)}_{+}$.
The simple ($N{=}\,1$) SUSY GCA will be defined as minimal supersymmetrization
which contains bosonic sector $h^{(d)}= O(2,1){\oplus}\,O(d)$; for the extended  ($N{>}\,1$)
SUSY GCA we should add also to  the bosonic sector of the superalgebra $G^{(d)}_{+}$
the Galilean internal sector (e.g. for $d{=}\,3$ it will be given by $U(N;\mathbb{H})\cong USp(2N)$).

The plan of our paper is the following. In Sect.\,2 we shall consider simple $N{=}\,1$ $d{=}3$ SUSY GCA
obtained as the suitable enlargement of the simplest supersymmetrization of the algebra
$O(2,1){\oplus}\,O(3)=O^\ast(4)=U_\alpha(2;{\rm \mathbb{H}})$ \footnote{
The quaternionic group (\ref{quatern}) can be obtained as the intersection of two complex
groups $OSp(4|2;{\rm \mathbb{C}})\cap SU(4;2)$.
For the description
of quaternionic algebras and superalgebras see e.g. \cite{KuTo,NovLuk,GKLuk}.
}
\begin{equation}\label{quatern}
U_\alpha(2|1;{\rm \mathbb{H}}) =OSp(4^\ast|\,2)
\end{equation}
which contain the bosonic sector $U_\alpha(2;{\rm \mathbb{H}}){\oplus}\,U(1;{\rm \mathbb{H}})\cong (O(2,1){\oplus}\,O(3)){}\oplus O(3)$.\footnote{
We recall that $U_\alpha(N;{\rm \mathbb{H}})$ describes the antiunitary quaternionic algebra which can be represented by
quaternionic $N{\times}N$ matrix generators ${\mathscr{U}}$ satisfying the relation
\begin{equation}\label{quat-1}
{\mathscr{U}}\, \Omega=-\Omega\,({\mathscr{U}})^T
\end{equation}
where the quaternionic  $N{\times}N$ antiHermitian metric $\Omega$ satisfies the condition
\begin{equation}\label{quat-2}
\Omega^+\equiv(\bar\Omega)^T=-\Omega\,.
\end{equation}
{}For even $N$ ($N{=}2k$) the matrix $\Omega$ can be chosen as antisymmetric Jordan matrix
$\left(
   \begin{array}{cc}
     0 & 1_k \\
     -1_k & 0 \\
   \end{array}
 \right)
$,
and for $N$ odd $\Omega$ is purely quaternionic-imaginary (e.g. for $N{=}1$ one can choose $\Omega=e_2$).
In complex notation we obtain that $U_\alpha(N)=O^\ast(2N)$. Further $U(N;{\rm \mathbb{H}})$
describes the unitary quaternionic algebra; in complex notation we get $U(N;{\rm \mathbb{H}})=
U(2N;{\rm \mathbb{H}}){\cap}Sp(2N;{\rm \mathbb{C}})\equiv USp(2N)$. Analogously, we get for quaternionic unitary superalgebra
\begin{equation}\label{quat-3}
U_\alpha U(M|N;{\rm \mathbb{H}})\,\,\cong\,\, SU(2M|2N) \cap OSp(2M|2N;{\rm \mathbb{C}})\,.
\end{equation}
The superalgebra $U_\alpha U(M|N;{\rm \mathbb{H}})$ for even $N$ is equivalent to the quaternionic-valued
superalgebra $OSp(N|M;{\rm \mathbb{H}})$~\cite{KuTo,NovLuk}.
}
It will be shown that the possibility of supersymmetrizing the Abelian subalgebra  ${\mathscr{A}}^{(9)}$
selects out from infinite family of superalgebras $D(2,1;\alpha)$ describing
versions of ${\cal N}{=}\,4$ superconformal mechanics \cite {IKL2,AIPT,MS,IL,FIL1,FIL2} only the choice $\alpha{=}\,1$, providing
$D(2,1;\alpha) \cong OSp\,(4^\ast|\,2)$.
In Sect.\,3 we shall consider the $N$-extended $d=3$ SUSY GCA and discuss the quaternionic structure
of $d=3$ Galilean symmetries and supersymmetries. We shall link the IW contraction procedure with
the standard nonrelativistic contraction described by $c\,{\rightarrow}\,\infty$ limit.
In Sect.\,4 we shall consider $N$-extended SUSY GCA for $d=1,2,4,5$.
In our final Sect.\,5 we provide an outlook, in particular we comment on the link
of $d$-dimensional SUSY GCA with ${\cal N}$-extended superconformal mechanics
and mention about the generalizations of SUSY GCA to the description of nonrelativistic $p$-branes
for $p{\geq}\,1$ (see also \cite{GKTown,GGKam,Sa}).

We add that there was proposed other supersymmetric extension of GCA (see \cite{BagM}),
outside our algebraic scheme, without supersymmetrization of all GCA generators.
In \cite{BagM} the GCA generators belonging to the $O(2,1)\,{\oplus}\,O(d)$
are not expressed as bilinears of supercharges. Because the Hamiltonian $\mathbf{H}$
belongs to $O(2,1)$, the supersymmetrization in  \cite{BagM} does not include the basic relation
$\{\mathbf{Q},\mathbf{Q}\}\,{\sim}\,\mathbf{H}$ of supersymmetric quantum mechanics,
what we evaluate as an argument in favour of our scheme.

\setcounter{equation}{0}
\section{$N{=}\,1$ $d{=}\,3$ Galilean superconformal algebra from the extension of $D(2,1;1)\,{\cong}\,OSp\,(4^\ast|2)$}

\quad\, Let us consider firstly the physical example of $d{=}\,3$, with the GCA described by the algebra
\begin{equation}\label{GCA-3}
g^{(3)}= \left( O(2,1)\oplus O(3) \right)\subset\!\!\!\!\!\!+ \,{\mathscr{A}}^{(9)}= h^{(3)}\subset\!\!\!\!\!\!+ \, k^{(3)}\,.
\end{equation}

The simplest supersymmetrization of the semisimple part $h^{(3)}=O(2,1)\oplus O(3)$ is provided by $OSp(3|2)\equiv OSp(3|2;\mathbb{R})$, but
it can be shown that in such a case it is not possible to enlarge such superalgebra by graded Abelian sector containing
${\mathscr{A}}^{(9)}=(\mathbf{P}_i, \mathbf{B}_i, \mathbf{F}_i)$ ($i=1,2,3$),
where $\mathbf{P}_i$ generate space translations, $\mathbf{B}_i$ yield
Galilean boosts and $\mathbf{F}_i$ produce the constant nonrelativistic accelerations. The next candidate for the supersymmetrization
of $h^{(3)}$ is $OSp(4|2)\equiv OSp(4|2;\mathbb{R})$, with internal sector $O(4)=O(3)\oplus O(3)$. In such a way in the process
of supersymmetrization of $O(2,1)\oplus O(3)$ we add still additional factor $O(3)$. Such bosonic symmetry describes
the symmetries of ${\cal N}{=}\,4$ superconformal mechanics. Interestingly enough,
we were also not able to show that the enlargement of superalgebra $OSp(4|2)$
permits the supersymmetrization of $g^{(3)}$ (see (\ref{GCA-3})) in a way described in previous section. Subsequently we
shall consider the one-parameter generalization $D(2,1;\alpha)$ of $OSp(4|2)$ superalgebra, where  $D(2,1;-\frac12)\cong OSp(4|2)$.
We shall show that the Abelian enlargement of $D(2,1;\alpha)$ leading to the supersymmetrizatrion of $g^{(3)}$ is possible
only if $\alpha=\,1$, in which case
\begin{equation}\label{D21-1}
D(2,1;1)\,\,\cong\,\, U_\alpha U(2;1|{\rm \mathbb{H}})\,\,\cong\,\,OSp\,(4^\ast|\,2)\,.
\end{equation}

We start from the set of the generators written in spinorial basis
\begin{equation}\label{D21-gen}
G^{(3)}_{+}=\left( \mathbf{Q}^{+}_{a \alpha A};\mathbf{R}_{ab}, \mathbf{J}_{\alpha\beta}, \mathbf{T}_{AB}\right)\,,
\end{equation}
which form the $D(2,1;\alpha)$ superalgebra (for details see for example \cite{Sorba,BILS,FIL2})
\begin{equation}\label{D21-QQ}
\{\mathbf{Q}^{+}_{a \alpha A}, \mathbf{Q}^{+}_{b \beta B}\}=
2\,\Big(\epsilon_{\alpha\beta}\epsilon_{AB} \mathbf{R}_{ab}+
\alpha\,\epsilon_{ab}\epsilon_{AB} \mathbf{J}_{\alpha\beta}-
(1+\alpha)\,\epsilon_{ab}\epsilon_{\alpha\beta} \mathbf{T}_{AB}\Big)\,,
\end{equation}
\begin{eqnarray}\label{D21-BB}
[\mathbf{R}_{ab}, \mathbf{R}_{cd}] &=&
i\left(\epsilon_{ac}\mathbf{R}_{bd} +\epsilon_{bd}\mathbf{R}_{ac}\right)\,,\\
{}[\mathbf{J}_{\alpha\beta}, \mathbf{J}_{\gamma\delta}] &=&
i\left(\epsilon_{\alpha\gamma}\mathbf{J}_{\beta\delta} +\epsilon_{\beta\delta}\mathbf{J}_{\alpha\gamma}\right)\,,\label{D21-BB-b}\\
{}[\mathbf{T}_{AB}, \mathbf{T}_{CD}] &=&
i\big(\epsilon_{AC}\mathbf{T}_{BD} +\epsilon_{BD}\mathbf{T}_{AC}\big)\,,\label{D21-BB-c}
\end{eqnarray}
\begin{eqnarray}\label{D21-BQ}
[\mathbf{R}_{ab}, \mathbf{Q}^{+}_{c \alpha A}] &=&
-i\,\epsilon_{c(a}\mathbf{Q}^{+}_{b) \alpha A}\,, \\
{}[\mathbf{J}_{\alpha\beta}, \mathbf{Q}^{+}_{a \gamma A}] &=&
-i\,\epsilon_{\gamma(\alpha}\mathbf{Q}^{+}_{a\beta)A}\,,\label{D21-BQ-b}\\
{}[\mathbf{T}_{AB}, \mathbf{Q}^{+}_{a \alpha C}] &=&
-i\,\epsilon_{C (A}\mathbf{Q}^{+}_{a \alpha B)}\,,\label{D21-BQ-c}
\end{eqnarray}
with other commutators vanishing.
All  $D(2,1;\alpha)$ generators are Hermitian, i.e. they satisfy the relations
\begin{equation}\label{Her-Q}
(\mathbf{Q}^{+}_{a \alpha A}){}^\dag = \epsilon^{\alpha\beta}\epsilon^{AB}\mathbf{Q}^{+}_{a \beta B}
\end{equation}
\begin{equation}\label{Her-B}
(\mathbf{R}_{ab})^\dag = \mathbf{R}_{ab}\,,\qquad
(\mathbf{J}_{\alpha\beta})^\dag = \epsilon^{\alpha\gamma}\epsilon^{\beta\delta}\mathbf{J}_{\gamma\delta}\,,\qquad
(\mathbf{T}_{AB}){}^\dag = \epsilon^{AC}\epsilon^{BD}\mathbf{T}_{CD}\,.
\end{equation}
Thus, the generators $\mathbf{R}_{ab}$ form the $O(2,1)$ algebra
($\mathbf{R}_{11}=\mathbf{H}$, $\mathbf{R}_{22}=\mathbf{K}$, $\mathbf{R}_{12}=\mathbf{D}$), whereas $\mathbf{J}_{\alpha\beta}$ and $\mathbf{T}_{AB}$
describe two $O(3)$ algebras which form $O(4)$ algebra which includes $d{=}\,3$ space rotation $\mathbf{J}_{\alpha\beta}$ and
$O(3)$ internal symmetry algebra $\mathbf{T}_{AB}$.
Indices $a,b,c=1,2$ are spinor $O(2,1)$ indices and  $\alpha,\beta,\gamma=1,2$;  $A,B,C=1,2$ are spinor indices of two $O(3)$ groups.
Everywhere in this paper we take $\epsilon_{12}=\epsilon^{21}=1$.
The link between $O(3)$ and $O(2,1)$ vectors and symmetric second-rank spinors is provided by corresponding $\sigma$-matrices:
$\mathbf{J}_{\alpha}{}^{\beta}=\mathbf{J}_{i}(\sigma_i)_{\alpha}{}^{\beta}$,
$\mathbf{R}_{a}{}^{b}=\mathbf{R}_{r}(\rho^r)_{a}{}^{b}$ and
$\mathbf{J}_{\alpha\beta}=\epsilon_{\beta\gamma}\mathbf{J}_{\alpha}{}^{\gamma}$,
$\mathbf{R}_{ab}=\epsilon_{ac}\mathbf{R}_{a}{}^{c}$.
The fermionic supercharges $\mathbf{Q}^{+}_{a \alpha A}$ unify in one $O(2,1)$ spinor the standard supercharges
$\mathbf{Q}^{+}_{\alpha A}=\mathbf{Q}^{+}_{1 \alpha A}$ and the generators of conformal supertranslations
$\mathbf{S}^{+}_{\alpha A}=\mathbf{Q}^{+}_{2 \alpha A}$.

Let us consider the graded Abelian enlargement of the $D(2,1;\alpha)$ superalgebra.
We add the following generators
\begin{equation}\label{D21-gen-centr}
G^{(3)}_{-}=\left( \mathbf{Q}^{-}_{a \alpha A}; \mathbf{A}_{ab,\alpha\beta}, \mathbf{A}_0\right)\,,
\end{equation}
where three 3-vectors $\mathbf{A}_{r,i}\cong({\mathbf{P}}_i, {\mathbf{B}}_i, {\mathbf{F}}_i)$ are
described in spinorial notation as $\mathbf{A}_{ab,\alpha\beta}$. The set of generators (\ref{D21-gen-centr}) forms the Abelian subalgebra
\begin{equation}\label{ab-centr}
\{\mathbf{Q}^{-}_{a \alpha A}, \mathbf{Q}^{-}_{b \beta B}\}=
[\mathbf{Q}^{-}_{a \alpha A}, \mathbf{A}_{bc,\beta\gamma}]=
[\mathbf{Q}^{-}_{a \alpha A}, \mathbf{A}_0]=[\mathbf{A}_{ab,\alpha\beta}, \mathbf{A}_{cd,\gamma\delta}]=
[\mathbf{A}_{ab,\alpha\beta}, \mathbf{A}_0]=0\,.
\end{equation}
The crossed (anti)commutators between $G^{(3)}_{-}$ and $G^{(3)}_{+}=D(2,1;\alpha)$
describe the following covariance relations
\begin{equation}\label{ab-D21-QQ}
\{\mathbf{Q}^{+}_{a \alpha A}, \mathbf{Q}^{-}_{b \beta B}\}=
\beta\,\epsilon_{AB}\mathbf{A}_{ab,\alpha\beta}+\gamma\,\epsilon_{ab}\epsilon_{\alpha\beta}\epsilon_{AB} \mathbf{A}_0 \,,
\end{equation}
\begin{equation}\label{ab-D21-QA}
[\mathbf{Q}^{+}_{a \alpha A}, \mathbf{A}_{bc,\beta\gamma}]=
-4i\,\epsilon_{a(b}\epsilon_{\alpha(\beta}\mathbf{Q}^{-}_{c)\gamma)A} \,,
\qquad
[\mathbf{Q}^{+}_{a ii^\prime}, \mathbf{A}_0]=
i\, \mathbf{Q}^{-}_{a ii^\prime}\,,
\end{equation}
\begin{eqnarray}
[\mathbf{R}_{ab}, \mathbf{Q}^{-}_{c \alpha A}] &=&
-i\,\epsilon_{c(a}\mathbf{Q}^{-}_{b) \alpha A}\,,\nonumber\\
{}[\mathbf{J}_{\alpha\beta}, \mathbf{Q}^{-}_{a \gamma A}] &=&
-i\,\epsilon_{\gamma(\alpha}\mathbf{Q}^{-}_{a\beta)A}\,,\label{ab-D21-BQ}\\
{}[\mathbf{T}_{AB}, \mathbf{Q}^{-}_{a \alpha C}] &=&
-i\,\epsilon_{C (A}\mathbf{Q}^{-}_{a \alpha B)}\,,\nonumber
\end{eqnarray}
\begin{equation} \label{ab-D21-BB}
\begin{array}{rcl}
[\mathbf{R}_{ab}, \mathbf{A}_{cd,\alpha\beta}]&=&-i\,\epsilon_{c(a}\mathbf{A}_{b)d, \alpha\beta}-i\,\epsilon_{d(a}\mathbf{A}_{b)c, \alpha\beta}\,,
\\[4pt]
[\mathbf{J}_{\alpha\beta}, \mathbf{A}_{ab,\gamma\delta}] &=&
-i\,\epsilon_{\gamma(\alpha}\mathbf{A}_{ab, \beta)\delta}-i\,\epsilon_{\delta(\alpha}\mathbf{A}_{ab, \beta)\delta}\,,
\end{array}
\end{equation}
\begin{equation} \label{ab-D21-BB0}
[\mathbf{T}_{AB}, \mathbf{A}_{ab,\alpha\beta}]=0\,,\qquad [\mathbf{R}_{ab}, \mathbf{A}_0]=
[\mathbf{J}_{\alpha\beta}, \mathbf{A}_0]=
[\mathbf{T}_{AB}, \mathbf{A}_0]=0\,.
\end{equation}
Reality properties of the generators (\ref{D21-gen-centr}) are described by
\begin{equation}\label{D21-Her-ab}
(\mathbf{Q}^{-}_{a \alpha A}){}^\dag = \epsilon^{\alpha\beta}\epsilon^{AB}\mathbf{Q}^{-}_{a \beta B}\,,
\qquad
(\mathbf{A}_{ab,\alpha\beta})^\dag = \epsilon^{\alpha\gamma}\epsilon^{\beta\delta}\mathbf{A}_{ab,\gamma\delta}\,,
\qquad
(\mathbf{A}_0)^\dag = \mathbf{A}_0\,.
\end{equation}

The commutators (\ref{ab-D21-BQ})-(\ref{ab-D21-BB0}) represent the standard form of $SO(2,1)$ and $SO(3)$ representations,
realized on corresponding spinor indices. On the right hand of the commutators (\ref{ab-D21-QQ}) we left
undetermined two numerical coefficients which remain not determined after all possible redefinitions of the generators $\mathbf{A}_{ab,\alpha\beta}$ and $\mathbf{A}_0$.

Let us check the consistency of the $D(2,1;\alpha)$ superalgebra (\ref{D21-QQ})-(\ref{D21-BQ-c}) and the relations (\ref{ab-centr})-(\ref{ab-D21-BB0}).
Already the Jacobi identities $\left(\mathbf{Q}^{+},\mathbf{Q}^{+},\mathbf{Q}^{-}\right)$
leads to unique values of the constants $\beta$ and $\gamma$ in (\ref{ab-D21-QQ})
\begin{equation} \label{be-ga}
\beta=\gamma=1\,.
\end{equation}
Moreover, these identities fix the value $\alpha= 1$ of the parameter $\alpha$ characterizing the $D(2,1;\alpha)$ superalgebra.
All other Jacobi identities are satisfied if $\alpha{=}\,1$ and the relations (\ref{be-ga}) are valid.

The $D(2,1;\alpha{=}1)$ superalgebra defines the $OSp(4^\ast|\,2)$ superalgebra (see \cite{Sorba,BILS,FIL2}).\footnote{
We can exchange the role of two $SU(2)$ algebras described by the generators $\mathbf{J}_{\alpha\beta}$ and $\mathbf{T}_{AB}$ in
(\ref{D21-QQ}) and consider other Abelian enlargement by the operators $\mathbf{A}_{ab,AB}$ in (\ref{ab-D21-QQ}).
Then, we shall obtain the condition $-(1+\alpha)=1$, i.e. $\alpha=-2$.
But $D(2,1;\alpha{=}1)\cong D(2,1;\alpha{=}{-}2) \cong OSp(4^\ast|\,2)$ \cite{Sorba,FIL2} and we obtain the same superalgebra.
}
Thus, we obtain that the $d{=}\,3$ SUSY GCA is given by the semidirect sum of the ${\rm OSp}(4^\ast|\,2)$
superalgebra $G^{(3)}_{+}$ (see (\ref{D21-gen}))
and the graded Abelian superalgebra (\ref{D21-gen-centr}).

We see that the superextension of $d{=}\,3$ GCA selects uniquely out of one-parameter family of $D(2,1;\alpha)$
supersymmetries only one representative
(\ref{D21-1}), which is endowed with quaternionic structure. It appears that in $d{=}\,3$ such quaternionic structure
has its origin in quaternionic structure of nonrelativistic $d{=}\,3$ $SU(2)$ spinors, due to the relation
$U(1;{\rm \mathbb{H}})\cong SU(2)$.\footnote{
If we describe nonrelativistic spinors by a pair of complex variables (nonrelativistic counterpart of Weyl spinors) the unitarity condition leads rather
to the group $U(2)$ instead of $SU(2)$.
}

The $N$-extended $d{=}\,3$ SUSY GCA will be derived in next Section as the result of the IW contraction of $2N$-extended relativistic
$D{=}3{+}1$ superconformal algebra $SU(2,2|2N)$ (see also \cite{Sa}) and compared with the derivation by
physical contraction $c\,{\rightarrow}\,\infty$ given in \cite{AL}. The quaternionic structure of  $d{=}\,3$
SUSY GCA implies, that one should contract the $D{=}\,4$ relativistic complex conformal superalgebra to the
superalgebra with the quaternionic structure.

\setcounter{equation}{0}
\section{$N$-extension of $d\,{=}\,3$ Galilean superconformal algebra and quaternionic structure}

We established in Sect.\,2 that the $N{=}\,1$ supersymmetrization of semisimple part of $d{=}\,3$
GCA is given by the semisimple superalgebra (\ref{D21-1}). In order to obtain the $N$-extended supersymmetrization we
extend the formula  (\ref{D21-1}) as follows:
\begin{equation}\label{D21-N}
G^{(3)}_{+}\,\,=\,\, U_\alpha U(2;N|{\rm \mathbb{H}})\,\,=\,\,OSp\,(4^\ast|\,2N)
\end{equation}
with the following basis
\begin{equation}\label{D21-gen-N}
G^{(3)}_{+}=\left\{ \mathbf{Q}^{+}_{a \alpha A}; \mathbf{R}_{ab}, \mathbf{J}_{\alpha\beta}, \mathbf{T}^{+ A}_{\ \,B}\right\}\,,
\end{equation}
where the $O^\ast(4)$ generators $\mathbf{R}_{ab}$, $\mathbf{J}_{\alpha\beta}$ describe in
spinorial notation the quantum-mechanical conformal algebra ${\rm O}(2,1)$ \cite{AFF,IKL} and
${\rm O}(3)$ space rotations; the generators $\mathbf{T}^{+ A}_{\ \,B}$ ($A,B=1,...,2N$) describe
the quaternionic internal algebra $U(N|{\rm \mathbb{H}})\cong USp(2N)$.
The $8N$ supercharges $\mathbf{Q}^{+}_{a \alpha A}$ are complex and satisfy
the quaternionic (or Majorana-symplectic) reality condition (see e.g. \cite{KuTo}).
In order to obtain the SUSY GCA we can decompose the $2N$-extended relativistic conformal superalgebra
in the following way
\begin{equation}\label{decom-3N}
SU(2,2|2N) \,\,=\,\, OSp(4^\ast|2N)\subset\!\!\!\!\!\!+ \, \frac{SU(2,2|2N)}{OSp(4^\ast|2N)} \,\,=\,\,
{\mathrm{H}}^{(3)}_N\subset\!\!\!\!\!\!+ \, \mathrm{K}^{(3)}_N
\end{equation}
where ${\mathrm{H}}^{(3)}_N{=}\,OSp\,(4^\ast|\,2N)$ and the IW contraction is obtained if we rescale the supercoset generators
\begin{equation}\label{WI-3N}
{\mathrm{H}}^{(3)}_N=\hat{\mathrm{H}}^{(3)}_N\,,\qquad {\mathrm{K}}^{(3)}_N=\lambda\,\hat{\mathrm{K}}^{(3)}_N;
\qquad\qquad \hat{\mathrm{H}}^{(3)}_N=G^{(3)}_{+}\,,\qquad \hat{\mathrm{K}}^{(3)}_N=G^{(3)}_{-}
\end{equation}
and perform the limit $\lambda\to\infty$.

Let us recall that $D=4$ $2N$-extended relativistic conformal algebra is obtained by adding to the
superPoincar\'{e} Weyl supercharges \footnote{
We define $D=4$ sigma-matrices as follows: $(\sigma^\mu)_{\alpha\dot\alpha}=(1_2,\vec{\sigma})_{\alpha\dot\alpha}$,
$(\tilde\sigma^\mu)^{\dot\alpha\alpha}=\epsilon^{\alpha\beta}\epsilon^{\dot\alpha\dot\beta}(\sigma^\mu)_{\beta\dot\beta}=
(1_2,-\vec{\sigma})^{\dot\alpha\alpha}$, $\sigma^{\mu\nu}=i\,\sigma^{[\mu}\tilde\sigma^{\nu]}$,
$\tilde\sigma^{\mu\nu}=i\,\tilde\sigma^{[\mu}\sigma^{\nu]}$. Always in this paper we use weight coefficient in (anti)symmetrization:
$A_{(\mu}B_{\nu)}=\frac12\,(A_{\mu}B_{\nu}+A_{\nu}B_{\mu})$, $A_{[\mu}B_{\nu]}=\frac12\,(A_{\mu}B_{\nu}-A_{\nu}B_{\mu})$.
The $D=4$ metric tensor is taken as $\eta_{\mu\nu}={\rm diag}(+1,-1,-1,-1)$. }
$Q_{\alpha}^{\,A}$, $\bar{Q}_{\dot{\alpha}\, A} =(Q_{\alpha}^{\,A})^\dagger$ ($A=1,...,2N$) satisfying the following basic SUSY relations
\begin{equation}\label{QQ-224}
\{Q_{\alpha}^{\,A}, \bar{Q}_{\dot{\beta} B} \} =
2(\sigma^\mu)_{\alpha \dot{\beta}} \, P_\mu  \,
\delta_{B}^{\, A}\,,
\qquad\quad \{Q_{\alpha}^{\,A}, {Q}_{\beta}^{\,B} \}
= \{\bar{Q}_{\dot{\alpha} A}, \bar{Q}_{\dot{\beta} B} \}=0 \,,
\end{equation}
the $2N$ conformal Weyl supercharges $S_{\alpha A}$, $\bar{S}_{\dot\alpha}^{\,A}= (S_{\alpha A})^\dagger$, supersymmetrizing the
conformal momenta $K_\mu$
\begin{equation}\label{SS-224}
\{ S_{\alpha A}, \bar{S}_{\dot{\beta}}^{\, B} \}
= 2(\sigma^\mu)_{\alpha \dot{\beta}} \, K_\mu  \,
\delta_{A}^{\, B}\,,\qquad\quad
\{ S_{\alpha A}, {S}_{{\beta} B} \}
=
\{ \bar{S}_{\dot{\alpha}}^{\, A},
\bar{S}_{\dot{\beta}}^{\, B} \} = 0 \,.
\end{equation}
Besides we have the relations
\begin{equation}\label{QS-224}
\begin{array}{rcl}
\{ Q_{\alpha}^{\,A}, S^{\,\beta}_{B} \}
&=&
-\delta^{A}_{B} (\sigma^{\mu\nu})_{\alpha}{}^{\beta}\, M_{\mu\nu} - 4i \,\delta_{\alpha}^{\beta} \, T^{\,A}_{B}
-2i\, \delta_{\alpha}^{\beta}\delta^{A}_{B} \left( D +i\,A\right)  \,,
\\[5pt]
\{ \bar{Q}_{\dot{\alpha} A},\bar{S}^{\,\dot{\beta} B} \}
&=&
-\delta^{B}_{A}(\tilde{\sigma}^{\mu\nu})^{\dot{\beta}}{}_{\dot{\alpha}}\, M_{\mu\nu} - 4i\, \delta^{\dot\beta}_{\dot\alpha} \, T^{\,B}_{\,A}
+2i\, \delta^{\dot\beta}_{\dot\alpha}\delta^{B}_{A} \left( D -i\,A\right) \,,
\\[5pt]
\{  Q_{\alpha}^{\,A},\bar{S}_{\dot{\beta}}^{\, B} \}
&=&
\{ \bar{Q}_{\dot{\alpha}A},{S}_{{\beta} B} \} = 0\,,
\end{array}
\end{equation}
Under the D=4 conformal generators ($M_{\mu\nu}, P_\mu, K _\mu, D$) the supercharges $Q^{i}_{\alpha}, S_{\alpha i}$ transform as follows
\begin{equation}\label{QM-224}
\begin{array}{lll}
&[M_{\mu\nu}, Q_{\alpha}^{\,A}] = - {\textstyle\frac{1}{2}}\, (\sigma_{\mu\nu})_{\alpha}{}^{\beta} Q_{\beta}^{\,A}\,,\qquad &
[M_{\mu\nu}, \bar{Q}_{\dot{\alpha} A}] =  {\textstyle\frac{1}{2}}\, (\tilde{\sigma}_{\mu\nu})^{\dot{\beta}}{}_{\dot{\alpha}} \, \bar{Q}_{\dot{\beta} A}\,,
\\[5pt]
&[M_{\mu\nu}, S_{\alpha A}] = - {\textstyle\frac{1}{2}}\, (\sigma_{\mu\nu})_{\alpha}{}^{\beta} S_{\beta A}\,,\qquad &
[M_{\mu\nu}, \bar{S}_{\dot{\alpha}}^{\, A}] = {\textstyle\frac{1}{2}}\,(\tilde{\sigma}_{\mu\nu})^{\dot{\beta}}{}_{\dot{\alpha}} \, \bar{S}_{\dot{\beta}}^{\, A}\,,
\end{array}
\end{equation}
\begin{equation}
[ P_\mu , Q_{\alpha}^{\,A}] = [ P_\mu , \bar{Q}_{\dot{\alpha} A}] =0\,,\qquad
[ P_\mu , S_{\alpha A}] =   ({\sigma}_\mu)_{{\alpha}\dot{\beta}} \, \bar{Q}_{\,A}^{\dot{\beta}}\,,\qquad
[ P_\mu , \bar{S}^{A}_{\dot{\alpha}}] =  - ({\sigma}_\mu)_{\beta\dot{\alpha}} \, {Q}^{\beta A} \,,
\end{equation}
\begin{equation}
[ K_\mu , Q_{\alpha}^{\, A}] =   ({\sigma}_\mu)_{{\alpha}\dot{\beta}} \, \bar{S}^{\dot{\beta A}}\,,\qquad
[ K_\mu , \bar{Q}_{\dot{\alpha}A}] =  - ({\sigma}_\mu)_{\beta\dot{\alpha}} \, {S}^{\beta}_{\,A} \,,\qquad
[ K_\mu , S_{\alpha A}] = [K_\mu , \bar{S}_{\dot{\alpha}}^{\, A}] =0\,,
\end{equation}
\begin{equation}
[ D , Q_{\alpha}^{\, A}] = {\textstyle\frac{i}{2}}\,  Q_{\alpha}^{\, A}\,,\quad
[ D , \bar{Q}_{\dot{\alpha}A}] =  {\textstyle\frac{i}{2}}\,\bar{Q}_{\dot{\alpha}A} \,,\qquad
[ D , S_{\alpha A}] = -{\textstyle\frac{i}{2}}\,S_{\alpha A}\,, \quad
[D , \bar{S}_{\dot{\alpha}}^{\, A}] =-{\textstyle\frac{i}{2}}\,\bar{S}_{\dot{\alpha}}^{\, A}\,.
\end{equation}
The generators $T^{A}_{B}$ describe $SU(2N)$ algebra
\begin{equation}\label{suN}
[T^{A}_{B},T^{C}_{D}]=i\left(\delta^{A}_{D}\,T^{C}_{B}-\delta^{C}_{B}\,T^{A}_{D}\right)  \,,\qquad
T^{A}_{A}=0\,,\qquad (T^{A}_{B})^\dagger = -T^{B}_{A}
\end{equation}
and transform supercharges as follows
\begin{equation}\label{I-Q}
\begin{array}{lll}
&[ T^{A}_{B}, Q_{\alpha}^{\, C} ] = \delta^{C}_{B} \, Q_{\alpha}^{\, A}
- {\textstyle\frac{1}{2N}} \, \delta^{A}_{B}\,  Q_{\alpha}^{\, C}\,,
\qquad &
[T^{A}_{B}, \bar{Q}_{\dot{\alpha}C} ] = -\delta^{A}_{C} \,\bar{Q}_{\dot{\alpha}B}
+ {\textstyle\frac{1}{2N}} \, \delta^{A}_{B}\,  \bar{Q}_{\dot{\alpha}C}\,,
\\[5pt]
&[T^{A}_{B}, {S}_{{\alpha}C} ] = -\delta^{A}_{C} \,{S}_{{\alpha}B}
+ {\textstyle\frac{1}{2N}} \, \delta^{A}_{B}\,  {S}_{{\alpha}C}\,,
\qquad &
[ T^{A}_{B}, \bar S_{\dot\alpha}^{\, C} ] = \delta^{C}_{B} \, \bar S_{\dot\alpha}^{\, A}
- {\textstyle\frac{1}{2N}} \, \delta^{A}_{B}\,  \bar S_{\dot\alpha}^{\, C}\,.
\end{array}
\end{equation}
The $U(1)$ axial charge $A=A^\dagger$ satisfies the relations
\begin{equation}\label{A-QS}
[ A , Q_{\alpha}^{A}] = -{\textstyle\frac{2-N}{2N}}\,  Q_{\alpha}^{A}\,,\quad
[ A , \bar{Q}_{\dot{\alpha}A}] =  {\textstyle\frac{2-N}{2N}}\,\bar{Q}_{\dot{\alpha}A} \,,\quad
[ A , S_{\alpha A}] = {\textstyle\frac{2-N}{2N}}\,S_{\alpha A}\,, \quad
[A , \bar{S}_{\dot{\alpha}}^{A}] =-{\textstyle\frac{2-N}{2N}}\,\bar{S}_{\dot{\alpha}}^{\, A}\,,
\end{equation}
and commutes with all other bosonic generators forming $O(4,2)\oplus SU(2N)$ algebra.
We see from (\ref{A-QS}) that for $N=2$ i.e. for $SU(2,2|4)$ the axial charge $A$ becomes a scalar central charge.

We rewrite the D=4 relativistic superconformal algebra in different fermionic Weyl basis \footnote{
We mention that the projections  (\ref{Q-pm}), (\ref{S-pm}) were introduced in \cite{AL}
for real Majorana supercharges.
}
\begin{eqnarray}\label{Q-pm}
& Q^{\pm A}_{\alpha} =
{\textstyle\frac{1}{\sqrt{2}}}\,\Big(Q_{\alpha}^{A} \pm\epsilon_{\alpha \beta}\,\Omega^{AB}\, \bar{Q}_{\dot{\beta}B}\,\Big)\,,
\qquad
& \bar{Q}^{\pm }_{\dot{\alpha}A}={\textstyle\frac{1}{\sqrt{2}}}\,\Big(\bar{Q}_{\dot{\alpha} A} \pm \epsilon_{\dot{\alpha} \dot{\beta}}\,
\Omega_{AB} \, {Q}^{B}_{\beta} \,\Big)\,,\\
\label{S-pm}
& S^{\pm }_{\alpha A} =
{\textstyle\frac{1}{\sqrt{2}}}\,\Big(S_{\alpha A} \mp\epsilon_{\alpha \beta}\,\Omega_{AB}\, \bar{S}_{\dot{\beta}}^{B}\,\Big)\,,
\qquad
& \bar{S}^{\pm A}_{\dot{\alpha}}={\textstyle\frac{1}{\sqrt{2}}}\,\Big(\bar{S}_{\dot{\alpha}}^{A} \mp \epsilon_{\dot{\alpha} \dot{\beta}}\,
\Omega^{AB}  {S}_{\beta B} \,\Big)\,,
\end{eqnarray}
where the real matrix $\Omega=(\Omega^{AB})$ ($\Omega_{AB}\equiv (\overline{\Omega^{AB}}) = \Omega^{AB})$ is a $2N\times 2N$ symplectic metric
($\Omega^2= -1$, $\Omega^T=-\Omega$) (see also eq.~(\ref{quat-2}) in footnote\,5)).
The relations (\ref{Q-pm}), (\ref{S-pm}) break Lorentz symmetry $O(3,1)$ to $O(3)$, and the
internal symmetry $U(2N)$ to $U(N|\mathbb{H})\cong USp(2N)$.

The Weyl supercharges (\ref{Q-pm}), (\ref{S-pm}) satisfy the subsidiary symplectic-Majorana conditions
\begin{equation}\label{MW-cond}
({Q}^{\pm A}_{\ {\alpha}})^\dagger
=\bar{Q}^{\pm}_{\dot{\alpha}A}
= \pm\, \epsilon_{\dot\alpha\dot\beta}
\Omega_{AB}\, Q^{\pm B}_{\beta} \,,\qquad\quad
({S}^{\pm}_{{\alpha} A})^\dagger =\bar{S}^{\pm A}_{\dot{\alpha}} = \mp\, \epsilon_{\dot\alpha\dot\beta} \, \Omega^{AB} \, S^{\pm }_{\, \beta B} \,.
\end{equation}
The subsidiary conditions (\ref{MW-cond}) describing quaternionic structure permits to choose as independent fermionic charges
Hermitian basis ($Q^{\pm A}_{\beta}$, $\bar{Q}^{\pm}_{\ \dot{\alpha}A}$,  $S^{\pm }_{\, \beta A}$, $\bar{S}^{\pm A}_{\dot{\alpha}}$;
$A=1,...,N$). Using holomorphic basis ($Q^{\pm A}_{\alpha}$, $S^{\pm }_{\, \alpha A}$; $A=1,...,2N$)
the reality condition (\ref{MW-cond}) allows to preserve properly
the covariance with respect to internal $USp(2N)$ symmetry.

The nonvanishing relations (\ref{QQ-224}), (\ref{SS-224}) can be represented in holomorphic basis as follows
\begin{eqnarray}\label{QQ-pm}
& \{ Q^{\pm A}_{\alpha} , Q^{\pm B}_{\beta} \}=\pm\,2\,
\Omega^{AB} \, \epsilon_{{\alpha} {\beta}} P_0 \,, \qquad &
\{ Q^{+ A}_{\alpha}, Q^{- B}_{\beta} \}=2\,
\Omega^{AB} \,  (\sigma_i )_{{\alpha} {\beta}}P_i \,\,,
\\[2pt]
\label{SS-pm}
& \{ {S}^{\pm}_{{\alpha}A},  {S}^{\pm}_{{\beta}B} \} =\pm\,2\,
\Omega_{AB} \, \epsilon_{{\alpha} {\beta}} K_0\,, \qquad &
\{ {S}^{+}_{{\alpha}A},  {S}^{-}_{{\beta}B}  \}
=-2\,\Omega_{AB} \,  (\sigma_i )_{{\alpha} {\beta}}K_i \,,
\end{eqnarray}
where $(\sigma_i )_{\alpha\beta}=(\sigma_i )_{\beta\alpha}=\epsilon_{\alpha\gamma}(\sigma_i )_{\beta\dot\gamma}$
or equivalently, using Hermitian basis of fermionic charges,
\begin{eqnarray}\label{QQ-pm-H}
& \{ Q^{\pm A}_{\alpha} , \bar{Q}^{\pm}_{\dot{\beta}B} \}=2\,
\delta^{A}_{B} \, \delta_{{\alpha} \dot{\beta}} P_0 \,, \qquad &
\{ Q^{+ A}_{\alpha}, \bar{Q}^{-}_{\dot{\beta}B} \}=2\,
\delta^{A}_{B} \,  (\sigma_i )_{{\alpha} \dot{\beta}}P_i \,\,,
\\[2pt]
\label{SS-pm-H}
& \{ {S}^{\pm}_{{\alpha}A},  \bar{S}^{\pm B}_{\dot{\beta}} \} =
-2\,\delta^{B}_{A} \, \delta_{{\alpha} \dot{\beta}} K_0\,, \qquad &
\{ {S}^{+}_{{\alpha}A},  \bar{S}^{- B}_{\dot{\beta}}  \}
=-2\,\delta^{B}_{A} \,  (\sigma_i )_{{\alpha} \dot{\beta}}K_i \,.
\end{eqnarray}
From (\ref{QS-224}) one obtains the following nonzero anticommutators
\begin{eqnarray}\label{QS-pp}
\{Q^{\pm A}_{\alpha}, S^{\pm}_{\beta B} \} &=& -\, \delta^A_{B}
\Big[ \epsilon_{ijk}(\sigma_{k})_{\alpha \beta}\, M_{ij} - 2i
\epsilon_{\alpha \beta} D \Big] + 2\,i\,\epsilon_{\alpha \beta}\, T^{+ A}_{\ \, B}\,, \\[3pt]
\label{QS-pm}
\{Q^{\pm A}_{\alpha}, S^{\mp}_{\beta B} \} &=& -2\, \delta^A_{B}
\Big[ i(\sigma_{i})_{\alpha \beta}\, M_{i0} + \epsilon_{\alpha \beta}\, A \Big]
+ 2\,i\,\epsilon_{\alpha \beta}\, T^{- A}_{\ \, B}\,,
\end{eqnarray}
where operators
\begin{equation}\label{T-pm}
T^{\pm A}_{\ \, B}\equiv T^A_B \pm \Omega^{AC}\, T^D_C\, \Omega_{DB}
\end{equation}
satisfy the relations
\begin{equation}\label{rel-T-pm}
T^{+ C}_{\ \,A}\Omega_{BC}=2\, T^{C}_{(A}\, \Omega_{B)C}\,,\quad
T^{- C}_{\ \,A}\Omega_{BC}=2\, T^{C}_{[A}\, \Omega_{B]C}\,;
\qquad
T^{\pm A}_{\ \,B}=\pm\, \Omega^{AC}T^{\pm D}_{\ \,C}\, \Omega_{DB}\,.
\end{equation}
Further one can also add the covariance relations (\ref{QM-224})--(\ref{A-QS})
written for supercharges (\ref{Q-pm}), (\ref{S-pm}).

The internal symmetry generators $T^{A}_{B}$ do split in relations (\ref{T-pm}) as follows
\begin{equation}\label{mosluke3.22}
\tilde h^{(3)}_N = \left(\,T^{+ A}_{\ \,B}\,\right) = \textit{USp}(2N)\,,
\qquad \qquad
\tilde k^{(3)}_N = \left(\,T^{- A}_{\ \,B}\,\right)= \frac{\textit{SU}(2N)}{\textit{USp}(2N)}\,,
\end{equation}
and provide an example of symmetric Riemannian space ($\tilde h^{(3)}_N, \tilde k^{(3)}_N$) with the algebraic relations
\begin{equation}\label{h-k}
[\tilde h^{(3)}_N, \tilde h^{(3)}_N] \subset \tilde h^{(3)}_N\,, \qquad
[\tilde h^{(3)}_N, \tilde k^{(3)}_N] \subset \tilde k^{(3)}_N\,, \qquad
[\tilde k^{(3)}_N, \tilde k^{(3)}_N] \subset \tilde h^{(3)}_N\,.
\end{equation}

In order to obtain the Galilean conformal superalgebra one can introduce the physical rescaling of the
relativistic supercharges \cite{AL}
\begin{equation}\label{QS-tr}
Q^{+ A}_{\alpha} = c^{-1/2} \, \mathbf{Q}^{+ A}_{\alpha}\,,
\qquad Q^{- A}_{\alpha} = c^{1/2} \, \mathbf{Q}^{- A}_{\alpha}\,,\qquad
S^{+}_{\alpha A} = c^{1/2} \, \mathbf{S}^{+}_{\alpha A}\,,
\qquad S^{-}_{\alpha A} = c^{3/2}\,\mathbf{S}^{-}_{\alpha A} \,.
\end{equation}
The physical rescaling of the bosonic generators of $O(4,2)\oplus USp(2k)$,
where $(P_\mu,M_{\mu\nu},D, K_\mu)\in O(4,2)$ and $T^{\pm i}_{\ \, j}\in U(2N)$, is the following
\begin{equation}\label{resc-o42}
\begin{array}{lll}
&& P_{0} = c^{-1}\,\mathbf{H}\,, \qquad P_{i} = \mathbf{P}_{i}\,,\qquad M_{ij}=  \mathbf{J}_{ij}\,,\qquad  M_{i0}=  c\,\mathbf{B}_{i}\,,
\qquad D=  \mathbf{D}\,,
\\[5pt]
&& K_{0} = c\,\mathbf{K}\,, \qquad  K_{i}=  c^2\,\mathbf{F}_{i}\,,\qquad
A = c\, \mathbf{A}_0\,, \qquad T^{+ A}_{\ \, B} = \mathbf{T}^{+ A}_{\ \, B}\,,
\qquad T^{- A}_{\ \, B} = c\,\mathbf{T}^{- A}_{\ \, B} \,.
\end{array}
\end{equation}
As a result of the contraction $c\to\infty$ we obtain in Weyl basis the $N$-extended SUSY GCA which was presented
in real Majorana basis for supercharges in \cite{AL}.

The physical rescaling (\ref{QS-tr})-(\ref{resc-o42}) is however not unique  because
the relativistic superalgebra $SU(2,2|2N)$ is invariant under the following rescaling:
\begin{equation}\label{resc-o224}
 P_{\mu}^\prime = \lambda\,P_{\mu}\,, \qquad  K_{\mu}^\prime = \lambda^{-1}\,K_{\mu}\,,
\qquad
Q^{\pm A\,\prime}_{\alpha} = \lambda^{1/2}\,Q^{\pm A}_{\alpha}\,, \qquad  S^{\pm\,\prime}_{\alpha A} = \lambda^{-1/2}\,S^{\pm}_{\alpha A}\,.
\end{equation}
If we compose rescaling (\ref{resc-o224}) with physical rescaling  (\ref{QS-tr})-(\ref{resc-o42})
we obtain
\begin{eqnarray}\label{resc-o224-compose}
\!\!\!\!\!\!\!\!\!\!& Q^{+ A \,\prime}_{\alpha} = \left(\frac{\lambda}{c}\right)^{\frac12}  \mathbf{Q}^{+ A}_{\alpha}\,,
\quad Q^{- A\,\prime}_{\alpha} = (\lambda c)^{\frac12} \, \mathbf{Q}^{- A}_{\alpha}\,,\qquad
S^{+\,\prime}_{\alpha A} = \left(\frac{c}{\lambda}\right)^{\frac12} \mathbf{S}^{+}_{\alpha A}\,,
\quad S^{-\,\prime}_{\alpha A} = \left(\frac{c^3}{\lambda}\right)^{\frac12}\mathbf{S}^{-}_{\alpha A}\,,
\nonumber \\[2pt]
\!\!\!\!\!\!\!\!\!\!&P^\prime_{0} = \frac{\lambda}{c}\,\mathbf{H}\,, \qquad P^\prime_{i} = \lambda\,\mathbf{P}_{i}\,,
\qquad M^\prime_{ij}=  \mathbf{J}_{ij}\,,\qquad  M^\prime_{i0}=  c\,\mathbf{B}_{i}\,,
\qquad D=  \mathbf{D}\,,
\\[2pt]
\!\!\!\!\!\!\!\!\!\!& K^\prime_{0} = \frac{c}{\lambda}\,\mathbf{K}\,, \qquad  K^\prime_{i}=  \frac{c^2}{\lambda}\,\mathbf{F}_{i}\,,\qquad
A^\prime = c\, \mathbf{A}_0\,, \qquad T^{+ A\,\prime}_{\ \, B} = \mathbf{T}^{+ A}_{\ \, B}\,,
\qquad T^{- A\,\prime}_{\ \, B} = c\,\mathbf{T}^{- A}_{\ \, B}\,.
\nonumber
\end{eqnarray}

If we put in (\ref{resc-o224-compose}) that $\lambda=c$ one can check that we obtain rescaling  (\ref{WI-3N})
of the supercoset decomposition of $SU(2,2|2N)$, with invariant subalgebra $OSp\,(4^\ast|\,2N)$,
spanned by generators $\mathbf{Q}^{+ A}_{\alpha}$, $\bar{\mathbf{Q}}^{+ }_{\dot\alpha A}$,
$\mathbf{S}^{+}_{\alpha A}$, $\bar{\mathbf{S}}^{+ A}_{\dot\alpha}$, $(\mathbf{H},\mathbf{K},\mathbf{D})\in O(2,1)$,
$\mathbf{J}_{ij}\in O(3)$, $\mathbf{T}^{+ A}_{\ \, B}\in USp\,(2N)$.
As a result of the contraction $\lambda=c\to \infty $ we obtain the following
fermionic bilinear relations of the $N$-extended SUSY GCA, with $8N$ complex supercharges
\begin{equation}\label{GCA-N-QQ}
\begin{array}{lc}
\!\!\!&\{ \mathbf{Q}^{+ A}_{\alpha}, \bar{\mathbf{Q}}^{+ }_{\dot\beta B} \}
= 2\,\delta_{\alpha \dot\beta} \,\delta^{A}_{B} \,  \mathbf{H}\,, \qquad
\{ \mathbf{S}^{+}_{{\alpha}A},  \bar{\mathbf{S}}^{+ B}_{\dot{\beta}} \} =-2\,\delta_{\alpha \dot\beta} \,\delta^{B}_{A} \,  \mathbf{K}\,,
\\[7pt]
\!\!\!&\{\mathbf{Q}^{+ A}_{\alpha}, \mathbf{S}^{+}_{\beta B} \} = -\, \delta^A_{B}
\Big[ \epsilon_{ijk}(\sigma_{k})_{\alpha \beta}\, \mathbf{J}_{ij} - 2i
\epsilon_{\alpha \beta} \mathbf{D} \Big] + 2\,i\,\epsilon_{\alpha \beta}\, \mathbf{T}^{+ A}_{\ \, B}\,,
\\[7pt]
\!\!\!&\{ \mathbf{Q}^{+ A}_{\alpha}, \bar{\mathbf{Q}}^{- }_{\dot\beta B} \}=
2\,\delta^{A}_{B} \,  (\sigma_i )_{{\alpha} {\dot\beta}}\mathbf{P}_i\,,\qquad
\{ \mathbf{S}^{+}_{{\alpha}A},  \bar{\mathbf{S}}^{- B}_{\dot{\beta}}  \}
=-2\,\delta^{B}_{A} \,  (\sigma_i )_{{\alpha} \dot{\beta}}\mathbf{F}_i\,,
\\[7pt]
\!\!\!&\{\mathbf{Q}^{\pm A}_{\alpha}, \mathbf{S}^{\mp}_{\beta B} \} = -2\, \delta^A_{B}
\Big[ i(\sigma_{i})_{\alpha \beta}\, \mathbf{B}_{i} + \epsilon_{\alpha \beta}\, \mathbf{A}_0 \Big]
+ 2\,i\,\epsilon_{\alpha \beta}\, \mathbf{T}^{- A}_{\ \, B}\,,
\\[7pt]
\!\!\!&\{ \mathbf{Q}^{- A}_{\alpha}, \mathbf{Q}^{- B}_{\beta} \}= 0\,,\qquad
\{ \mathbf{S}^{-}_{{\alpha}A},  \mathbf{S}^{-}_{{\beta}B}  \} = 0\,,\qquad
\{ \mathbf{Q}^{- A}_{\alpha}, \mathbf{S}^{-}_{{\beta}B} \} = 0 \,.
\end{array}
\end{equation}
The relations in mixed bosonic-fermionic sector are the following
\begin{equation}\label{GCA-N-HQ}
\begin{array}{lc}
\!\!\!\!\!\!&[ \mathbf{H}, \mathbf{S}^{\pm}_{{\alpha}A}] = \bar{\mathbf{Q}}^{\pm \dot\alpha}_{\ A}\,,
\quad [ \mathbf{K}, \mathbf{Q}^{\pm A}_{\alpha}] = \bar{\mathbf{S}}^{\pm \dot\alpha  A}\,,\qquad
[\mathbf{D}, \mathbf{Q}^{\pm A}_{\alpha} ] =  {\textstyle\frac{i}{2}}\, \mathbf{Q}^{\pm A}_{\alpha}\,,
\quad [\mathbf{D}, \mathbf{S}^{\pm}_{{\alpha}A} ] = -{\textstyle\frac{i}{2}}\, \mathbf{S}^{\pm}_{{\alpha}A}\,,
\\[7pt]
\!\!\!\!\!\!&[\mathbf{J}_{ij}, \mathbf{Q}^{\pm A}_{\alpha}] = - {\textstyle\frac{1}{2}}\,\epsilon_{ijk} (\sigma_{k})_{\alpha}{}^{\beta} \,
\mathbf{Q}^{\pm A}_{\beta}\,, \qquad [\mathbf{J}_{ij}, \mathbf{S}^{\pm}_{{\alpha}A}]
= - {\textstyle\frac{1}{2}}\,\epsilon_{ijk} (\sigma_{k})_{\alpha}{}^{\beta} \, \mathbf{S}^{\pm}_{{\beta}A}\,,
\\[7pt]
\!\!\!\!\!\!&[\mathbf{P}_i, \mathbf{S}^{+}_{{\alpha}A}] = - (\sigma_i)_{\alpha\dot\beta} \, \bar{\mathbf{Q}}^{- \dot\beta}_{\ A}\,, \qquad
[\mathbf{F}_i, \mathbf{Q}^{+ A}_{\alpha}] = - (\sigma_i)_{\alpha\dot\beta} \, \bar{\mathbf{S}}^{-\dot{\beta} A}\,,
\\[7pt]
\!\!\!\!\!\!&[\mathbf{B}_{i}, \mathbf{Q}^{+ A}_{\alpha}] =  {\textstyle\frac{i}{2}}\, (\sigma_{i})_{\alpha}{}^{\beta} \,
\mathbf{Q}^{- A}_{\beta}\,, \qquad [\mathbf{B}_{i}, \mathbf{S}^{+}_{{\alpha}A}]
=  {\textstyle\frac{i}{2}}\, (\sigma_{i})_{\alpha}{}^{\beta} \, \mathbf{S}^{-}_{{\beta}A}\,,
\\[7pt]
\!\!\!\!\!\!&[\mathbf{A}_0, \mathbf{Q}^{+ A}_{\alpha} ] = {\textstyle\frac{1}{2}}\, (1 - {\textstyle\frac{2}{N}}\,)\, \mathbf{Q}^{- A}_{\alpha}\,,
\qquad [\mathbf{A}_0, \mathbf{S}^{+}_{{\alpha}A} ] = {\textstyle\frac{1}{2}}\, (1 - {\textstyle\frac{2}{N}}\,)\, \mathbf{S}^{-}_{{\alpha}A}\,.
\end{array}
\end{equation}

The covariance relations with respect to the generators $\hat{\tilde h}{}^{(3)}_N = (\,\mathbf{T}^{+ A}_{\ \,B}\,)$ and
$\hat{\tilde k}{}^{(3)}_N = (\,\mathbf{T}^{- A}_{\ \,B}\,)$ take the form
\begin{equation}\label{GCA-N-QT-pl}
[\mathbf{T}^{+ A}_{\ \, B}, \mathbf{Q}^{\pm C}_{\alpha} ] = \left(\mathscr{U}^A_B\right){}^C_D\, \mathbf{Q}^{\pm D}_{\alpha}\,,
\qquad [\mathbf{T}^{+ A}_{\ \, B}, \mathbf{S}^{\pm}_{{\alpha}C} ] = -\left(\mathscr{U}^A_B\right){}^D_C\, \mathbf{S}^{\pm}_{{\alpha}D}\,,
\end{equation}
\begin{equation}\label{GCA-N-QT-min}
\begin{array}{ll}
[\mathbf{T}^{- A}_{\ \, B}, \mathbf{Q}^{+ C}_{\alpha} ] = \left(\tau^A_B\right){}^C_D\, \mathbf{Q}^{- D}_{\alpha}\,,
\quad &[\mathbf{T}^{- A}_{\ \, B}, \mathbf{S}^{+}_{{\alpha}C} ] = -\left(\tau^A_B\right){}^D_C\, \mathbf{S}^{-}_{{\alpha}D}\,,
\\[5pt]
[\mathbf{T}^{- A}_{\ \, B}, \mathbf{Q}^{- C}_{\alpha} ] = 0\,,
\quad &[\mathbf{T}^{- A}_{\ \, B}, \mathbf{S}^{-}_{{\alpha}C} ] = 0\,,
\end{array}
\end{equation}
where the $2N{\times} 2N$ matrix $\left(\mathscr{U}^A_B\right)$
\begin{equation}\label{U-Usp}
\left(\mathscr{U}^A_B\right){}^C_D = \delta^A_D\delta^C_B -\Omega^{AC}\Omega_{BD}
\end{equation}
defines the representation of the $U\!Sp\,(2N)$ algebra and we define
\begin{equation}\label{tau-Usp}
\left(\tau^A_B\right){}^C_D = \delta^A_D\delta^C_B +\Omega^{AC}\Omega_{BD} -{\textstyle\frac{1}{N}}\,\delta^A_B\delta^C_D\,.
\end{equation}

In $N\,{=}\,1$ case the contracted algebra  (\ref{GCA-N-QQ})-(\ref{GCA-N-QT-min}) coincides
with the superalgebra considered in Sect.\,2. We note that in $N\,{=}\,1$ case, when
$\Omega_{AB}=\epsilon_{AB}$, $A=1,2$, we have $T^{- A}_{\ \, B}\equiv 0$ and $\left(\tau^A_B\right){}^C_D\equiv 0$
(see (\ref{T-pm}) and (\ref{tau-Usp})).

{}From (\ref{WI-3N}) follows that the coset generators (see (\ref{h-k})) are rescaled ($\tilde h^{(3)}_N=\hat{\tilde h}{}^{(3)}_N$,
${\tilde k}^{(3)}_N=\lambda\,\hat{\tilde k}{}^{(3)}_N$) and in the limit $\lambda\to\infty$ we get
\begin{equation}\label{h-k-til}
[\hat{\tilde h}{}^{(3)}_N, \hat{\tilde h}{}^{(3)}_N] \subset \hat{\tilde h}{}^{(3)}_N\,, \qquad
[\hat{\tilde h}{}^{(3)}_N, \hat{\tilde k}{}^{(3)}_N] \subset \hat{\tilde k}{}^{(3)}_N\,, \qquad
[\hat{\tilde k}{}^{(3)}_N, \hat{\tilde k}{}^{(3)}_N] =0\,.
\end{equation}
The generators $\hat{\tilde k}{}^{(3)}_N$ described in  (\ref{GCA-N-QQ}) by generators $\mathbf{T}^{- A}_{\ \,B}$
are Abelian and they describe a sort of internal complex momenta.
One gets the following structure of Galilean conformal internal symmetry algebra
\begin{equation}\label{mosluke3.27}
\hat{\mathbf{T}} = \hat {\tilde h}{}^{(3)}_N\,\subset\!\!\!\!\!\!+\, \hat {\tilde k}{}^{(3)}_N\,,
\qquad \qquad
\begin{array}{l}
\hat {\tilde h}{}^{(3)}_N= U(N|\mathbb{H})=\textit{USp}(2N)\,,
\\[2pt]
\hat {\tilde k}{}^{(3)}_N=\hat\mathscr{A}^{N(2N-1)} \quad \hbox{(Abelian)} \,,
\end{array}
\end{equation}
i.e. general structure of $G^{(3)}$ (see (\ref{S-gen}) is realized by the following Galilean generators
\begin{eqnarray}
G^{(3)}_{+} &=& \Big( \underbrace{\mathbf{Q}^{+ A}_{\alpha}, \bar{\mathbf{Q}}^{+}_{\dot\alpha A},
\mathbf{S}^{+}_{\alpha A}, \bar{\mathbf{S}}^{+ A}_{\dot\alpha}}_{\mathbf{Q}^{+}_{a\alpha A}}\quad;\quad
\underbrace{\mathbf{H}, \mathbf{K}, \mathbf{D}}_{\mathbf{R}_{ab}=O(2,1)}\quad ;
\underbrace{\mathbf{J}_{ij}}_{\mathbf{J}_{\alpha\beta}=O(3)};
\underbrace{\mathbf{T}^{+ A}_{\ \, B}}_{USp(2N)}
\Big)=OSp(4^\ast|2N)\,, \label{gen-3N+}\\
G^{(3)}_{-} &=& \Big( \underbrace{\mathbf{Q}^{- A}_{\alpha}, \bar{\mathbf{Q}}^{-}_{\dot\alpha A},
\mathbf{S}^{-}_{\alpha A}, \bar{\mathbf{S}}^{- A}_{\dot\alpha}}_{\mathbf{Q}^{-}_{a\alpha A}}\quad;\quad
\underbrace{\mathbf{P}_i, \mathbf{B}_i, \mathbf{F}_i}_{\mathbf{A}_{ab,\alpha\beta}}\quad;\quad
\mathbf{A}_{0}\quad;\quad
\mathbf{T}^{- A}_{\ \, B}\Big)\,.\label{gen-3N-}
\end{eqnarray}

The $N$-extended SUSY GCA has a quaternionic structure. {}For $N\,{=}\,1$ the compact internal symmetry is described by
$SU(2)\simeq U(1;\mathbb{H})$ and for arbitrary $N$ by
$USp(2N)\simeq U(N;\mathbb{H})$. The Galilean conformal supercharges
($\mathbf{Q}^{\pm A}_{\alpha}$, $\bar{\mathbf{Q}}^{\pm}_{\dot\alpha A}$,
$\mathbf{S}^{\pm}_{\alpha A}$, $\bar{\mathbf{S}}^{\pm A}_{\dot\alpha}$) can be described as the
$N$ quadruplets of quaternionic supercharges.
Indeed, the two-component Weyl spinor $Z_\alpha =(z_1,z_2)$ can be described by a single quaternion $q$ (see e.g. \cite{KuTo}) as follows
\begin{equation}\label{mosluke3.42}
z_1=x_1 + iy_1,\quad z_2 = x_2 + i y_2\qquad
\leftrightarrow\qquad q^\mathbb{H} = x_1 +e_3 y_1 +
e_2(x_2 + e_3 y_2) \,.
\end{equation}
We see that the $N$-extended $d\,{=}\,3$ SUSY GCA is generated by $N$ quadruplets $\mathbf{Q}^{\pm \mathbb{H}}_{M}$,
$\mathbf{S}^{\pm \mathbb{H}}_{M}$ of quaternionic supercharges,
where the indices $M$ span the quaternionic representation space of Galilean compact
internal symmetry $U(N;\mathbb{H})$ group. It
is a matter of quite tedious calculations (compare e.g. with \cite{KuTo}, Sect.~6) to reexpress
$N$-extended $d\,{=}\,3$ SUSY GCA in the quaternionic form.


\setcounter{equation}{0}
\section{$N$-extended superconformal mechanics and their links with
Galilean superconformal algebras for $d\,{=}\,1,2,4,5$}

\subsection{$d\,{=}\,1$}

\quad\, The set $G^{(1)}_{\,+}$ of real generators is the following ($a,b=1,2$)
\begin{equation}\label{osp12-gen}
G^{(1)}_{+}=\left( \mathbf{Q}^{+}_{a}; \mathbf{R}_{ab}\right)\,,\qquad
(\mathbf{Q}^{+}_{a}){}^\dag = \mathbf{Q}^{+}_{a}\,,
\qquad
(\mathbf{R}_{ab})^\dag = \mathbf{R}_{ab}= \mathbf{R}_{ba}\,.
\end{equation}
which form the real $OSp\,(1|2)$ superalgebra describing the graded symmetries of ${\cal N}{=}\,1$ superconformal
mechanics \cite{AIPT,MS}. The nonvanishing (anti)commutators are  (\ref{D21-BB}) and
\begin{equation}\label{osp12-QQ}
\{\mathbf{Q}^{+}_{a}, \mathbf{Q}^{+}_{b} \}=
2\, \mathbf{R}_{ab}\,,
\qquad
[\mathbf{R}_{ab}, \mathbf{Q}^{+}_{c}]=-i\,\epsilon_{c(a}\mathbf{Q}^{+}_{b)}\,.
\end{equation}

The graded Abelian enlargement of the $OSp\,(1|2)$ superalgebra is given by
generators
\begin{equation}\label{osp12-gen-centr}
G^{(1)}_{-}=\left( \mathbf{Q}^{-}_{a}; \mathbf{A}_{ab} \right),\qquad
(\mathbf{Q}^{-}_{a}){}^\dag = \mathbf{Q}^{-}_{a}\,,
\qquad
(\mathbf{A}_{ab})^\dag = \mathbf{A}_{ab}= \mathbf{A}_{ba}\,,
\end{equation}
which form the Abelian subalgebra
\begin{equation}\label{osp12-ab-centr1}
\{\mathbf{Q}^{-}_{a}, \mathbf{Q}^{-}_{b}\}=
[\mathbf{Q}^{-}_{a}, \mathbf{A}_{bc}]=
[\mathbf{A}_{ab}, \mathbf{A}_{cd}]= 0
\end{equation}
and transform as the $OSp\,(1|2)$ representation as follows
\begin{equation}\label{ab-osp12-QQ}
\{\mathbf{Q}^{+}_{a}, \mathbf{Q}^{-}_{a}\}=\beta\,\mathbf{A}_{ab}\,,
\end{equation}
\begin{equation}\label{ab-osp12-QA}
[\mathbf{Q}^{+}_{a}, \mathbf{A}_{bc}]=2i\,\epsilon_{a(b} \mathbf{Q}^{-}_{c)}\,,
\end{equation}
\begin{equation} \label{ab-osp12-BB}
[\mathbf{R}_{ab}, \mathbf{Q}^{-}_{c}]=-i\,\epsilon_{c(a}\mathbf{Q}^{-}_{b)}\,,
\qquad
[\mathbf{R}_{ab}, \mathbf{A}_{cd}]=-i\,\epsilon_{c(a}\mathbf{A}_{b)d}-i\,\epsilon_{d(a}\mathbf{A}_{b)c}
\end{equation}
where we postulated in  (\ref{ab-osp12-QQ}) that one real parameter $\beta$ is not determined.
It can be seen however  that
the Jacobi identities $\left(\mathbf{Q}^{+},\mathbf{Q}^{+},\mathbf{Q}^{-}\right)$
fix the value $\beta=1$ of the constant $\beta$. {}For such choice of $\beta$
all other Jacobi identities are satisfied.

The finite-dimensional $d{=}\,1$ relativistic conformal superalgebra with $D{=}\,2$
conformal algebra $O(2,1) {\oplus}\, O(2,1)=Sp(2) {\oplus}\, Sp(2)$ is provided
by the sum of the $d{=}\,0$ Galilean conformal superalgebras $OSp(1|2){\oplus}\, OSp(1|2)$.
The $d{=}\,1$ simple SUSY GCA can be obtained by IW contraction of supercoset
decomposition
\begin{equation} \label{osp12-decomp}
OSp(1|2){\oplus}\, OSp(1|2) = OSp(1|2)_D\subset\!\!\!\!\!\!+ \, \frac{OSp(1|2)_L{\oplus}\, OSp(1|2)_R}{OSp(1|2)_D}\,,
\end{equation}
where the generators $G^{(1)}_{+}= OSp(1|2)_D$ are obtained by taking the diagonal
sums $\frac{1}{\sqrt{2}}(\hat g_L +\hat g_R)=\hat g_D$ of left and right generators.
The contraction of rescaled difference $\frac{1}{\sqrt{2}}(\hat g_L -\hat g_R)$
provides $3$ generators $\mathbf{A}_{ab}$.

The family of extended $d{=}\,1$ SUSY GCA is parametrized by a pair of numbers $(N,M)$ and
can be defined as the suitable IW contractions
of the supercoset decompositions  of the superalgebras $OSp(N|2)_L{\oplus}\, OSp(M|2)_R$.

\subsection{$d\,{=}\,2$}

\quad\, We start from the following set $G^{(2)}_{+}$ of the generators
\begin{equation}\label{osp22-gen}
G^{(2)}_{+}=\left( \mathbf{Q}^{+}_{a}, \bar\mathbf{Q}^{+}_{a}; \mathbf{R}_{ab}, \mathbf{J}, \mathbf{C}\right)\,,
\end{equation}
which define the $SU(1,1|1)\,{\cong}\,OSp\,(2|2)$ superalgebra with one central charge (see e.e. \cite{AP,FR,IKL2,AIPT})
\begin{equation}\label{osp22-QQ}
\{\mathbf{Q}^{+}_{a}, \bar\mathbf{Q}^{+}_{b} \}=
2\left( \mathbf{R}_{ab}+ \epsilon_{ab} \mathbf{J}+ \epsilon_{ab} \mathbf{C}\right)\,,
\end{equation}
\begin{equation} \label{osp22-BQ1}
[\mathbf{R}_{ab}, \mathbf{Q}^{+}_{c}]=-i\,\epsilon_{c(a}\mathbf{Q}^{\,+}_{b)}\,,\qquad
[\mathbf{R}_{ab}, \bar\mathbf{Q}^{+}_{c}]=-i\,\epsilon_{c(a}\bar\mathbf{Q}^{\,+}_{b)}\,,
\end{equation}
\begin{equation} \label{osp22-BQ2}
[\mathbf{J}, \mathbf{Q}^{+}_{a}]=-{\textstyle\frac{i}{2}}\,\mathbf{Q}^{+}_{a}\,,\qquad
[\mathbf{J}, \bar\mathbf{Q}^{+}_{a}]={\textstyle\frac{i}{2}}\,\bar\mathbf{Q}^{+}_{a}\,,
\end{equation}
where the commutators (\ref{D21-BB}) and other (anti)commutators vanish.
Hermiticity properties of the  $SU(1,1|1)$ generators are the following
\begin{equation}\label{osp22-Her-Q}
(\mathbf{Q}^{+}_{a}){}^\dag = \bar\mathbf{Q}^{+}_{a}\,,
\qquad
(\mathbf{R}_{ab})^\dag = \mathbf{R}_{ab}= \mathbf{R}_{ba}\,,\qquad
(\mathbf{J})^\dag = -\mathbf{J}\,,\qquad
(\mathbf{C})^\dag = -\mathbf{C}\,.
\end{equation}
The generators $\mathbf{R}_{ab}$ form $SO(1,2)$ algebra, whereas $\mathbf{J}$
is the $O(2)$ generator of $d\,{=}\,2$ space rotations. The generator $\mathbf{C}$
provides the central charge.
Here we use the realization of the $OSp\,(2|2)$ algebra
as in \cite{IKL2} with fermionic supercharges being complex $O(2)$ spinors.

We propose now the enlargement of the $SU(1,1|1)$ superalgebra defining $d\,{=}\,2$ SUSY GCA.
We add the generators
\begin{equation}\label{osp22-gen-centr}
G^{(2)}_{-}=\left( \mathbf{Q}^{-}_{a}, \bar\mathbf{Q}^{-}_{a}; \mathbf{A}_{ab},\bar\mathbf{A}_{ab} \right),
\qquad (\mathbf{Q}^{-}_{a}){}^\dag = \bar\mathbf{Q}^{-}_{a}\,,
\qquad
(\mathbf{A}_{ab})^\dag = \bar\mathbf{A}_{ab}\,,
\qquad
\mathbf{A}_{ab}= \mathbf{A}_{ba}\,,
\end{equation}
which form graded Abelian subalgebra
\begin{equation}\label{osp22-ab-centr1}
\{\mathbf{Q}^{-}_{a}, \mathbf{Q}^{-}_{b}\}=\{\mathbf{Q}^{-}_{a}, \bar\mathbf{Q}^{-}_{b}\}=
[\mathbf{Q}^{-}_{a}, \mathbf{A}_{bc}]=[\mathbf{Q}^{-}_{a}, \bar\mathbf{A}_{bc}]=
[\mathbf{A}_{ab}, \mathbf{A}_{cd}]=[\mathbf{A}_{ab}, \bar\mathbf{A}_{cd}]= 0
\end{equation}
and transform as the following $SU(1,1|1)$ representations (see (\ref{S+-}))
\begin{equation}\label{ab-osp22-QQ1}
\{\mathbf{Q}^{+}_{a}, \mathbf{Q}^{-}_{b}\}=2\beta\,\mathbf{A}_{ab}\,,\qquad
\{\bar\mathbf{Q}^{+}_{a}, \bar\mathbf{Q}^{-}_{b}\}=2\beta\,\bar\mathbf{A}_{ab} \,,
\end{equation}
\begin{equation}\label{ab-osp22-QA}
[\mathbf{Q}^{+}_{a}, \bar\mathbf{A}_{bc}]=-2i\,\epsilon_{a(b}\bar\mathbf{Q}^{-}_{c)}\,,\qquad
[\bar\mathbf{Q}^{+}_{a}, \mathbf{A}_{bc}]=-2i\,\epsilon_{a(b}\mathbf{Q}^{-}_{c)} \,,
\end{equation}
\begin{equation}\label{ab-osp22-QA0}
[\mathbf{C},\mathbf{Q}^{-}_{a}]=-i\,\gamma\,\mathbf{Q}^{-}_{a}\,,\quad
[ \mathbf{C},\bar\mathbf{Q}^{-}_{a}]=i\,\gamma\, \bar\mathbf{Q}^{-}_{a} \,,
\qquad
[\mathbf{C},\mathbf{A}_{ab}]=-i\,\gamma\,\mathbf{A}_{ab}\,,\quad
[ \mathbf{C},\bar\mathbf{A}_{ab}]=i\,\gamma\, \bar\mathbf{A}_{ab} \,,
\end{equation}
\begin{equation} \label{ab-osp22-BQ1}
[\mathbf{T}_{ab}, \mathbf{Q}^{-}_{c}]=-i\,\epsilon_{c(a}\mathbf{Q}^{-}_{b)}\,,\qquad
[\mathbf{T}_{ab}, \bar\mathbf{Q}^{-}_{c}]=-i\,\epsilon_{c(a}\bar\mathbf{Q}^{-}_{b)}\,,
\end{equation}
\begin{equation} \label{ab-osp22-BB1}
[\mathbf{T}_{ab}, \mathbf{A}_{cd}]=-i\,\epsilon_{c(a}\mathbf{A}_{b)d}-i\,\epsilon_{d(a}\mathbf{A}_{b)c}
\,,\qquad [\mathbf{T}_{ab}, \bar\mathbf{A}_{cd}]=-i\,\epsilon_{c(a}\bar\mathbf{A}_{b)d}-i\,\epsilon_{d(a}\bar\mathbf{A}_{b)c}\,,
\end{equation}
\begin{equation} \label{ab-osp22-BB2}
[\mathbf{J}, \mathbf{Q}^{-}_{a}]=-{\textstyle\frac{i}{2}}\,\mathbf{Q}^{-}_{a}\,,\quad
[\mathbf{J}, \bar\mathbf{Q}^{-}_{a}]={\textstyle\frac{i}{2}}\,\bar\mathbf{Q}^{-}_{a}\,,\qquad [\mathbf{J}, \mathbf{A}_{ab}]=-i\,\mathbf{A}_{ab}
\,,\quad [\mathbf{J},\bar\mathbf{A}_{ab}]=i\,\bar\mathbf{A}_{ab}\,.
\end{equation}
We introduced in (\ref{ab-osp22-QQ1}), (\ref{ab-osp22-QA0}) two real parameters $\beta$ and $\gamma$.
If we check the consistency of the $SU(1,1|1)$ superalgebra (\ref{osp22-QQ})-(\ref{osp22-BQ2}) with
the relations (\ref{ab-osp22-QQ1})-(\ref{ab-osp22-BB2}) one can show that only
the Jacobi identities $\left(\mathbf{Q}^{+},\bar\mathbf{Q}^{+},\mathbf{Q}^{-}\right)$
leads to a complete fixing of the constants $\beta$ and $\gamma$ in (\ref{ab-osp22-QQ1}), (\ref{ab-osp22-QA0}), namely
$
\beta=\gamma=1
$.
All other Jacobi identities are then satisfied.

For consistency it is important the presence of $\mathbf{C}$ in the superalgebra $G^{(2)}_{+}$.
This generator is the central charge in the superalgebra  $G^{(2)}_{+}$, but in the
full superalgebra  $G^{(2)}$ it produces the $U(1)$ transformations acting on $G^{(2)}_{-}$ sector
(see (\ref{ab-osp22-QA0})).

{}For $N$-extended SUSY GCA as basic subsuperalgebra $G^{(2)}_{\,+}$ we take the central extension of the $SU(1,1|\,N)$ superalgebra,
\begin{equation}\label{S+-gen-2}
G^{(2)}_{+}=SU(1,1|\,N)\oplus\,U(1)\,,
\end{equation}
which can be introduced for any $N\geq 2$ (see  \cite{Ram}). Bosonic subalgebra is $SU(1,1)\,{\oplus}\,U(N)\,{\oplus}\,U(1)$;
the $U(1)$ factor in the decomposition $U(N)=SU(N)\,{\oplus}\,U(1)$ acts nontrivially on fermionic charges,
except if $N{=}\,2$ case when its describes a central charge \cite{IKL2}.

In order to obtain $N$-extended $d{=}\,2$ SUSY GCA we introduce
supercoset decomposition of $D{=}\,1{+}2$ relativistic conformal superalgebra
$OSp\,(4|\,2N)$
\begin{equation}\label{decomp-gen-2}
OSp\,(4|\,2N)=\Big(SU(1,1|\,N) \oplus\,U(1)\Big) \subset\!\!\!\!\!\!+ \, \frac{OSp\,(4|\,2N)}{SU(1,1|\,N)\oplus \,U(1)}
= \mathrm{H}^{(2)}_{N}\subset\!\!\!\!\!\!+ \,  \mathrm{K}^{(2)}_{N}\,.
\end{equation}
The factor $\mathrm{K}^{(2)}_{N}=\frac{OSp\,(2N|\,4)}{SU(1,1|\,N)\oplus\,U(1)}$ contains product of two bosonic factor
$\frac{Sp\,(4)}{SU(1,1){\oplus}U(1)}\,{\cdot}\frac{O(2N)}{U(N)}$. After IW contraction
$\mathrm{H}^{(2)}_{N}\,{\subset\!\!\!\!\!\!+ }\,\mathrm{K}^{(2)}_{N} \rightarrow G^{(2)}_{+}\,{\subset\!\!\!\!\!\!+ }\,G^{(2)}_{-}$ the first factor
$\frac{Sp\,(4)}{SU(1,1){\oplus}U(1)}$ produces the sets of the generators $(\mathbf{P}_i,\mathbf{B}_i,\mathbf{F}_i)$, $i=1,2$ and from
second factor $\frac{O(2N)}{U(N)}$ one gets internal bosonic Abelian charges of $N$-extended $d{=}\,2$ SUSY GCA.

\subsection{$d\,{=}\,4$}

In this case, the simple relativistic $D{=}\,5$ superconformal algebra is described by the exceptional superalgebra
$F(4)$  \cite{Ram}, or more precisely, by its real form given by the real superalgebra $F(4)=F(4;2)$ \cite{Sorba}, which contains the bosonic subalgebra
$O(5,2){\oplus}\,O(3)$. We stress that this choice is different from other real superalgebra
$F(4)=F(4;0)\ni O(2,1){\oplus}\,O(7)$ which has been used in ${\cal N}{=}\,8$ superconformal mechanics \cite{MS,DI}.
The $N{>}\,1$ $D{=}\,5$ superconformal algebras can be obtained (see \cite{FuOha,Ber})
only by the dimensional reduction of $D{=}\,6$ superconformal algebras, discussed below in Sect.\,4.4.

The supersymmetrization $G^{(4)}_{+}$ of the bosonic semisimple algebra $O(2,1)\,{\oplus}\,O(4)$ leads
as in $d{=}\,3$ case to the superalgebra $G^{(3)}_{+}= OSp\,(4^\ast|\,2)$
defined in  (\ref{D21-QQ})-(\ref{D21-BQ}).
If we introduce supercoset decomposition
\begin{equation}\label{decomp-gen-4}
F(4;2)=OSp\,(4^\ast|\,2)\subset\!\!\!\!\!\!+ \, \frac{F(4;2)}{OSp\,(4^\ast|\,2)}
=\mathrm{H}^{(4)}_{N}\subset\!\!\!\!\!\!+ \,\mathrm{K}^{(4)}_{N}
\end{equation}
and perform the IW  contraction
$\mathrm{H}^{(4)}_{N}\,{\subset\!\!\!\!\!\!+ }\,\mathrm{K}^{(4)}_{N} \rightarrow G^{(4)}_{+}\,{\subset\!\!\!\!\!\!+ }\,G^{(4)}_{-}$
we shall obtain $d{=}\,4$ SUSY GCA.
The supercoset $\mathrm{K}^{(4)}_{N}$ becomes graded Abelian superalgebra $G^{(4)}_{-}$
with 8 fermionic graded Abelian supercharges $\mathbf{Q}^{-}_{a \alpha A}$ and 15 bosonic Abelian ones, obtained from the contraction
of $\frac{O(5,2){\oplus}O(3)}{O(2,1){\oplus}O(4)}$, which include $d{=}\,4$ generators  $(\mathbf{P}_i,\mathbf{B}_i,\mathbf{F}_i)$.

\subsection{$d\,{=}\,5$}

In order to get $d{=}\,5$ SUSY GCA we can use again the method of IW contraction.
Firstly let us introduce $D{=}\,6$ superconformal algebra, which should contains as a factor
the $D{=}\,6$ conformal algebra $O(6,2)\cong U_\alpha(4;\mathbb{H})$. We arrive at the following
quaternionic $N$-extended $D{=}\,6$ superconformal algebra \cite{Ram,HLM,KuTo}
\begin{equation}\label{su-conf6}
U_\alpha U(4|N;\mathbb{H})\,\cong\, OSp\,(8^\ast|\,2N)
\end{equation}
with $16N$ real supercharges and bosonic sector $O(6,2)\,{\oplus}\,USp(2N)$.

In order to define the superalgebra $G^{(5)}_{+}$ one should supersymmetrize the bosonic sector
$g^{(5)}=O(2,1)\,{\oplus}\,O(5)$, where $O(2,1)\,{\cong}\,Sp(2)\,{\cong}\,SU(1,1)$ and $O(5)\,{\cong}\,U(2;\mathbb{H})\,{\cong}\,USp(4)$.
In supersymmetrization procedure we should therefore embed $O(2,1)$ algebra into an algebra with quaternionic structure.
Using $U_\alpha(2\,;\mathbb{H})=O(2,1)\,{\oplus}\,O(3)$ we arrive at the following minimal superalgebra $G^{(5)}_{+}$
\begin{equation}\label{s-d5}
G^{(5)}_{+}= U_\alpha U(2|2;\mathbb{H})\,\cong\, OSp\,(4^\ast|\,4)
\end{equation}
with $16$ real supercharges and bosonic subalgebra
$U_\alpha(2;\mathbb{H})\,{\oplus}\,U(2;\mathbb{H})\cong O(3)\,{\oplus}\,O(2,1)\,{\oplus}\,O(5)$.

The $N{=}\,1$ $d{=}\,5$ SUSY GCA is obtained by IW contraction of the following coset decomposition
of $D{=}\,6$ $N{=}\,2$ relativistic conformal superalgebra
\begin{equation}\label{decomp-5}
OSp\,(8^\ast|\,4)=OSp\,(4^\ast|\,4) \subset\!\!\!\!\!\!+ \, \frac{OSp\,(8^\ast|\,4)}{OSp\,(4^\ast|\,4) }
= \mathrm{H}^{(5)}\subset\!\!\!\!\!\!+ \,\mathrm{K}^{(5)}\,.
\end{equation}
After IW contraction
$\mathrm{H}^{(5)}\,{\subset\!\!\!\!\!\!+ }\,\mathrm{K}^{(5)} \rightarrow G^{(5)}_{+}\,{\subset\!\!\!\!\!\!+ }\,G^{(5)}_{-}$
the supercoset $\mathrm{K}^{(5)}$ becomes graded Abelian superalgebra $G^{(5)}_{-}$
with 16 fermionic graded Abelian charges and 22 bosonic Abelian ones which include $d{=}\,5$ generators  $(\mathbf{P}_i,\mathbf{B}_i,\mathbf{F}_i)$
obtained from the contraction of $\frac{O(6,2)}{O(2,1){\oplus}O(3)}$.

{}For $N{>}\,1$ one should consider the contraction of $D{=}\,6$ superconformal algebra (\ref{su-conf6})
with even $N=2n$. The extended $d{=}\,5$ superalgebra $G^{(5)}_{+}$ will take a form
\begin{equation}\label{s-gen-d5}
G^{(5)}_{+}= U_\alpha U(2|2n;\mathbb{H})\,\cong\, OSp\,(4^\ast|\,4n)
\end{equation}
and the $N$-extended $d{=}\,5$ SUSY GCA is given by the IW contraction of the following supercoset
\begin{equation}\label{decomp-gen-5}
OSp\,(8^\ast|\,4n)=OSp\,(4^\ast|\,4n) \subset\!\!\!\!\!\!+ \, \frac{OSp\,(8^\ast|\,4n)}{OSp\,(4^\ast|\,4n) }
= \mathrm{H}^{(5)}_{N}\,{\subset\!\!\!\!\!\!+ }\,\mathrm{K}^{(5)}_{N}\,.
\end{equation}
The graded superalgebra $G^{(5)}_{-}$ obtained from $\mathrm{K}^{(5)}_{N}$
contains $16n$ fermionic anticommuting charges and $22$ bosonic Abelian charges.
We see that the number of bosonic generators in the subalgebra $G^{(5)}_{-}$ does not depend on $n$.

\setcounter{equation}{0}
\section{Final remarks}

The aim of this paper is the description of explicit algebraic structures of SUSY GCA in $1\,{\leq}\,d\,{\leq}\,5$ space dimensions.
In the dimensions $d\,{=}\,1,2,3,5$ for $D\,{=}\,d+1$ space-time do exist $N$-extended relativistic conformal superalgebras with
arbitrary $N$ and we present also in these dimensions the $N$-extended SUSY GCA;
for $d\,{=}\,4$ exist only unique relativistic $D\,{=}\,5$ conformal superalgebra \cite{Ram,Sorba}
and corresponding $d\,{=}\,4$ SUSY GCA.
As it was discussed in detail for $d\,{=}\,3$ in Sect.\,2,
the SUSY GCA can be also obtained as an enlargement of semisimple superalgebra $G^{(d)}_{+}$
which supersymmetrizes the bosonic semisimple part $h^{(d)}$ of full SUSY GCA. We observe that our
choices of simple (nonextended) superalgebras $G^{(d)}_{+}$
are linked with the superalgebras describing the ${\cal N}$-extended supersymmetries in the models of superconformal mechanics.
In particular,
\begin{equation}
G^{(d)}_{+}= \left\{
\begin{array}{llcl}
OSp\,(1|2) \quad &\mbox{for}\quad d=1&\cong&\quad{\cal N}=1\,,\\
SU(1,1|\,1)\,{\oplus}\,U(1) \quad &\mbox{for}\quad d=2&\cong&\quad{\cal N}=2\,, \\
OSp\,(4^\ast|2) \quad &\mbox{for}\quad d=3\,\,\,\,\mbox{and}\,\,\,\,d=4\quad&\cong&\quad{\cal N}=4\,,\\
OSp\,(4^\ast|4) \quad &\mbox{for}\quad d=5&\cong&\quad{\cal N}=8\,.
\end{array}
\right.
\end{equation}
Note that in cases $d\,{=}\,3$ and $d\,{=}\,4$ the superalgebras $G^{(3)}_{+}$ and $G^{(4)}_{+}$
are the same, with the bosonic sector of $OSp\,(4^\ast|2)$ equal to $h^{(4)}=O(3)\,{\oplus}\, h^{(3)}$,
i.e. we see that the $d\,{=}\,3$ internal symmetry sector $O(3)$ is incorporated in $d\,{=}\,4$ into the space symmetry sector,
$O(4)=O(3)\,{\oplus}\, O(3)$.

An important task is to apply the present results for the models with a physical interpretation.
We mention here the following possibilities:
\\ \\
{\bf{i)}} In  \cite{FKLM} we introduced the fundamental spinorial realization of the subalgebra $h^{(3)}=O(2,1){\oplus}\,O(3)$
of $d\,{=}\,3$ GCA as introducing nonrelativistic counterpart of  $D\,{=}\,4$ relativistic twistors.
Analogously, the superalgebras $G^{(3)}_{+}$ can be used as the ones which define
by its fundamental graded representations the nonrelativistic counterpart of $D\,{=}\,4$ relativistic supertwistors~\cite{Fer}.
\\ \\
{\bf{ii)}} One can extend the coset techniques providing in \cite{FIL-W}
the geometric $\sigma$-models of GCA-invariant classical mechanics to the supersymmetric $\sigma$-models invariant under SUSY GCA.
In particular by keeping on SUSY Galilean conformal group manifold the
time $t$ and space coordinates $x_i$ as independent parameters
we can promote remaining SUSY GCA parameters to the $D\,{=}\,(d{+}1)$-dimensional nonrelativistic Goldstone fields.
Using for $d\,{=}\,3$ scalar four-forms as action densities \cite{Volkov} one can arrive at the
nonrelativistic field-theoretical models invariant under SUSY GCA.
\\ \\
{\bf{iii)}} In the known nonrelativistic versions of AdS/CFT correspondence \cite{Son,BalMcC}
the Schr\"{o}dinger (super)algebras \footnote{
{}For the description of superSchr\"{o}dinger symmetry see \cite{DuHo1,Sa1}.
}
are used as describing nonrelativistic (SUSY) CFT.
It is interesting to look for other nonrelativistic limit of AdS/CFT, with contracted AdS symmetries (see e.g. \cite{Gib})
corresponding to the Galilean nonrelativistic CFT with GC symmetries.
{}For $D\,{=}\,1{+}1$ ($d\,{=}\,1$) the relativistic CA as well as its nonrelativistic GCA limit are infinite-dimensional \cite{BagGop,Bag2}.
For $d\,{\geq}2$ the relativistic CA is finite-dimensional but in nonrelativistic theory there are there are the infinite-dimensional
versions of GCA describing the sets of conformal isometries of Galilean space-time  \cite{NOR,DuHo,Bag2}.
In such a context it is interesting to observe the $D\,{=}3$ correspondence \cite{Bag1} between the infinite-dimensional nonrelativistic
conformal isometries and Bondi-Metzner-Sachs group \cite{BMS} which describes asymptotic isometries of flat
Minkowski space at null infinity. We point out that it is interesting to consider the supersymmetrized versions
of all these conformal symmetry schemes.
\\ \\
{\bf{iv)}} In this paper we considered the rescaling (\ref{resc-o42}) of generators
corresponding to the redefinitions $\vec{x}^{\,\prime} \,{=}\,\vec{x}$, $x_0\,{=}\,c\,t$ of space and time coordinates before applying
the $c\,{\rightarrow}\,\infty$ contraction limit.
Such redefinition introducing nonrelativistic time $t$ is natural and well justified if we consider the classical mechanics models
with distinguished role of time as the evolution parameter. If we wish to study
the nonrelativistic contraction of the $p$-brane dynamics in $D\,{=}\,d{+}1$ space-time we should split the total space-time $(x_0,x_1,...,x_d)$ into
the $p\,{+}\,1$ coordinates $(x_0,x_1,...,x_p)$ describing the worldvolume of $p$-brane (inner ``generalized time'' manifold)
and the remaining outer coordinates transversal to $p$-brane. If we rescale differently the ``inner'' and ``outer'' coordinates
of (super)$p$-brane one obtains in the contraction limit
more general nonrelativistic (super)symmetries \cite{GKTown,GGKam,ADV,Sa}
described by the corresponding so-called semi-Galilean (super)algebras.
We can add that semi-Galilean CA are useful as well if we wish to generalize
the BMS/GCA correspondence to $D\,{>}\,3$  \cite{Bag1}.
\\

Many issues presented above are now under our consideration.

\section*{Acknowledgements}

\noindent We would like to thank Evgeny Ivanov for valuable comments.
We acknowledge a support from the grant of the Bogoliubov-Infeld Programme and RFBR
grants 09-02-01209, 09-01-93107 (S.F.), as well as
from the Polish Ministry of Science and Higher Educations grant No.~N202331139 (J.L.).
S.F. thanks the members of the Institute of Theoretical Physics at Wroclaw University
for warm hospitality.

\end{document}